\documentclass[a4paper,11pt]{article}
\pdfoutput=1 

\usepackage{jheppub,bm,mathtools,slashed} 
\usepackage[usenames,dvipsnames,svgnames,table,x11names]{xcolor}
\usepackage{extarrows}
\usepackage{easybmat}

\usepackage{nicematrix}

\usepackage[T1]{fontenc} 
\usepackage[utf8]{inputenc}

\renewcommand*{\backref}[1]{}
\renewcommand*{\backrefalt}[4]{%
  \ifcase #1 %
    \relax 
  \or
    {\scriptsize (page~#2).}%
  \else
    {\scriptsize (pages~#2).}%
  \fi%
}

\definecolor{light_blue}{rgb}{0.15, 0.35, 0.95}
\definecolor{kit_green}{rgb}{0, 
0.58823 
, 0.50980 
}

\usepackage{tikz}
\usepackage{pgfplots}
\pgfplotsset{compat=1.14}
\usepackage{tikz-3dplot}
\usetikzlibrary{intersections, calc,positioning,decorations.pathreplacing,decorations.pathmorphing,arrows,arrows.meta,patterns,pgfplots.fillbetween,math,matrix}

\usepackage{multirow}
\usepackage{booktabs} 
\usepackage{siunitx} 
\usepackage{tabularx} 

\usepackage{graphicx}
\usepackage{amsmath}
\usepackage{slashed}
\usepackage{amssymb}
\usepackage{amsfonts}
\usepackage{physics}
\usepackage{mathrsfs}
\usepackage{bm}
\usepackage{url}
\usepackage{ytableau}
\usepackage{amsthm}
\usepackage{tcolorbox}
\usepackage{ascmac}
\usepackage{tikz}
\usepackage{mathtools}

\usepackage{dynkin-diagrams}
\usetikzlibrary{backgrounds}
\usepackage{tikz-cd}

\newcommand{\C}{\mathbb{C}}

\newcommand{\hilb}{\mathcal{H}}

\newcommand{\R}{\mathbb{R}}

\newcommand{\Z}{\mathbb{Z}}
\newcommand{\Q}{\mathbb{Q}}

\newcommand{\Spin}{\mathrm{Spin}}
\newcommand{\SO}{\mathrm{SO}}

\newcommand{\SU}{\mathrm{SU}}
\newcommand{\U}{\mathrm{U}}

\newcommand{\Sp}{\mathrm{Sp}}

\newcommand{\g}{\mathfrak{g}}

\usepackage{cancel}

\newcommand{\ve}{\varepsilon}

\makeatletter
  
  \@addtoreset{equation}{section}

\title{
Investigating 9d/8d non-supersymmetric branes and theories from supersymmetric heterotic strings
}

\author[1,2]{Yuta Hamada}
\author[2]{and Arata Ishige}
\affiliation[1]{Theory Center, IPNS, High Energy Accelerator Research Organization (KEK), 1-1 Oho, Tsukuba, Ibaraki 305-0801, Japan}
\affiliation[2]{Graduate Institute for Advanced Studies, SOKENDAI, 1-1 Oho, Tsukuba, Ibaraki 305-0801, Japan}

\emailAdd{yhamada@post.kek.jp}
\emailAdd{arata@post.kek.jp}

\abstract{
We consider heterotic string theories in nine and eight dimensions. We identify the disconnected part of the spacetime gauge group by studying the outer automorphism of the charge lattices. The absence of the global symmetry indicates the existence of non-supersymmetric codimension two branes. Moreover, we provide a list of gauge groups and matter contents of non-supersymmetric rank-reduced heterotic string theories (a branch corresponding to the $E_8$ string on $S^1$) from the orbifolding of the outer automorphism as well as the fermion parity. We also provide examples in eight dimensions.
}

\preprint{KEK-TH-2652}

\begin{document}
\maketitle
\flushbottom

\section{Introduction}
Among various string theories, heterotic strings have long attracted phenomenological interest due to reasons such as possessing non-abelian gauge group symmetry even in the absence of branes. There are two supersymmetric heterotic string theories in 10d: $(E_8\times E_8)\rtimes\mathbb{Z}_2$ and $\Spin(32)/\Z_2$ theories. They correspond to the $E_8\times E_8$ root lattice and the $\Spin(32)/\Z_2$ lattice, respectively.
It is also important to study the non-supersymmetric string theories, as they could be relevant to the real world. See e.g. for \cite{Blaszczyk:2014qoa,Hamada:2015ria,Ashfaque:2015vta,Blaszczyk:2015zta,Itoyama:2019yst,Itoyama:2020ifw,Itoyama:2021itj} for studies in this direction. 
In fact, on top of the well-known 10d supersymmetric heterotic string theories, there exist many 10d non-supersymmetric heterotic theories. For instance, non-supersymmetric and non-tachyonic $\mathrm{O}(16)\times \mathrm{O}(16)$ theory is known~\cite{Dixon:1986iz,Alvarez-Gaume:1986ghj}.
The 10d non-supersymmetric string theories are listed in \cite{Kawai:1986vd}.
In addition to the $\mathrm{O}(16)\times \mathrm{O}(16)$ theory, there are $\SO(32), \mathrm{O}(16)\times E_8$, $\mathrm{O}(8)\times \mathrm{O}(24)$, $(E_7\times\SU(2))^2$, $\U(16)$ and $E_8$ theories.
The classification in 10d was recently revisited~\cite{BoyleSmith:2023xkd,Rayhaun:2023pgc,Hohn:2023auw} with a modern understanding of fermionization~\cite{Tachikawa:2018,Karch:2019lnn}.
All 10d non-supersymmetric heterotic string is constructed by $\mathbb{Z}_2$ orbifold of 10d supersymmetric heterotic string.
For instance, the 10d $E_8$ string is obtained via a twist that exchanges two $E_8$'s in $E_8\times E_8$ theory. Other non-supersymmetric strings are obtained by inner automorphism twists.

As a next step, it is natural to study the structure of 9d supersymmetric and non-supersymmetric heterotic strings. Contrary to 10d, 9d heterotic string has the Narain moduli space, and many kinds of gauge groups are realized by going through the moduli space. For the 9d/8d supersymmetric string, all possible charge lattices corresponding to maximal gauge enhancement are identified in \cite{Font:2020rsk} (see also \cite{Cachazo:2000ey}). The similar analysis for the rank reduced CHL string~\cite{Chaudhuri:1995fk} and 7d/6d theories is done in \cite{Font:2021uyw,Fraiman:2021soq,Fraiman:2021hma}.

\begin{figure}[t]
    \centering
\begin{tikzpicture}

\node at (-2, 4) {\textbf{10d}};
\node at (-2, 0) {\textbf{9d}};

\node[red] at (0, 5) {\textbf{SUSY}};
\node[blue] at (6, 5) {\textbf{non-SUSY}};

\node[red] at (-0.5, 4) {$\SO(32)$};
\node[red] at (1, 4) {$E_8 \times E_8$};

\node[red] (Rank17) at (-0.5, 0) {Rank 17};
\node[red] (CHL) at (1, 0) {CHL};
\node at (-0.2, 2) {$S^1$};
\node[magenta] at (1.2, 2) {RT};
\node[olive] at (2.9, 2) {RT$(-1)^F$};
\node[blue] at (3, 0) {$B_{IIb}$};
\draw[->, thick] (0.5, 3.5) -- (-0.25, 0.5);
\draw[->, thick] (-0.5, 3.5) -- (-0.5, 0.5);
\draw[->, thick, magenta] (0.7, 3.5) -- (1, 0.5);
\draw[->, thick, olive] (1.2, 3.5) -- (2.8, 0.5);

\node[blue] (E8) at (6, 4) {$E_8$};
\node[blue] at (8, 4) {$(E_7 \times \SU(2))^2$};
\node at (10, 4) {$\cdots$};

\node[blue] (BIII) at (6, 0) {$B_{III}$};
\draw[->, thick] (6, 3.5) -- (6, 0.5);
\node at (5.7, 2) {$S^1$};
\node at (9, 2) {$S^1$};
\node[brown] at (7.5, 2) {R};
\node[blue] at (7.7, 0) {$B_{IIa}$};
\node[blue] at (9, 0) {$A_I$};
\draw[->, thick, brown] (7.7, 3.5) -- (7.7, 0.5);
\draw[->, thick] (8, 3.5) -- (9, 0.5);
\draw[->, thick, green!50!black] (Rank17) to[out=-45, in=-135] (BIII);
\node[green!50!black] at (3, -1) {This work};

\end{tikzpicture}
    \caption{Relation among heterotic string theories in 10d and 9d. Here black arrow corresponds to simple $S^1$ compactifications.  The colored arrow involves the $\mathbb{Z}_2$ twist on the charge lattice (denoted by R), $(-1)^F$ twist, and half shift on $S^1$ (denoted by T). Regarding 9d non-supersymmetric theories, we have used notation in \cite{Hohn:2023auw,DeFreitas:2024ztt}. Note that $A_I$ is rank $17$ while $B_{IIa}$, $B_{IIb}$, and $B_{III}$ are rank $9$. The red and blue colors correspond to supersymmetric and non-supersymmetric theories, respectively.}
    \label{fig:theories}
\end{figure}
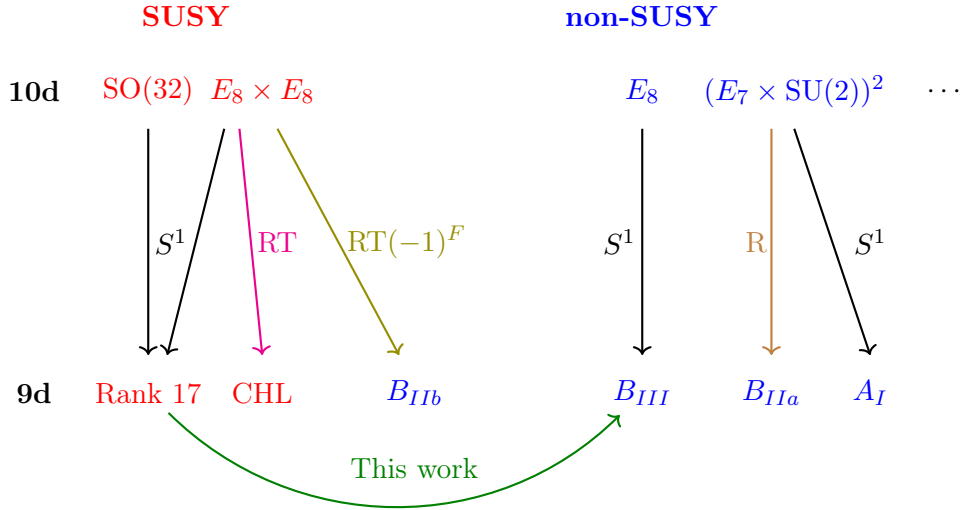

Regarding 9d non-supersymmetric heterotic string theories, it is argued that there are four disconnected branches at least perturbatively~\cite{Nakajima:2023zsh,DeFreitas:2024ztt}.\footnote{See also \cite{Baykara:2024tjr,Angelantonj:2024jtu,Detraux:2024esd} for recent studies on non-supersymmetric strings.} See Fig.~\ref{fig:theories} for the relation between 10d and 9d strings. One of the theory ($A_I$) has rank 17 while other threes ($B_{IIa}, B_{IIb}, B_{III}$) has a reduced rank 9. The maximal enhanced gauge group for $A_I$ is studied in \cite{Fraiman:2023cpa} while the other are not.

In this paper, we study the gauge groups and matter contents of rank reduced $B_{III}$ theory, which can be done as $\mathbb{Z}_2$ orbifolding of 9d supersymmetric heterotic strings. We also establish the relation between 9d supersymmetric and non-supersymmetric heterotic string theories. 
When the $\mathbb{Z}_2$ symmetry of the orbifold is an outer automorphism of the charge lattice, this indicates that the $D$-dimensional spacetime gauge group is disconnected. A disconnected gauge group contains dual $(D-2)$-form symmetry. The no global symmetry conjecture implies the existence of non-supersymmetric branes to break $(D-2)$-form symmetry.
In this way, we can enumerate non-supersymmetric branes in 9d supersymmetric heterotic string theory.
As a result, we obtain the following eight gauge symmetries:
$C_9,~C_8+ A_1,~C_6+ A_2+ A_1,~C_5+ A_4,~C_4+ D_5,~C_3+ E_6,~C_2+ E_7,~C_1+ E_8$.
We mainly focus on 9d, but we also provide several examples in 8d.

The organization of the paper is as follows. In Sec.~\ref{sec:orbifold}, we review the supersymmetric heterotic string and its orbifold to get non-supersymmetric strings. In Sec.~\ref{sec:fermionization}, we discuss the relation between supersymmetric and non-supersymmetric heterotic string theories. The non-supersymmetric branes from the disconnected gauge group are argued in Sec.~\ref{sec:branes}. In Sec.~\ref{sec:9d}, we study the 9d heterotic strings. In Sec.~\ref{sec:8d}, we provide several examples in 8d.
The Sec.~\ref{sec:future} is devoted to discussions and future directions.
The technical details are summarized App.~\ref{sec:theta} and App.~\ref{sec:lie algebra}.

\section{Asymmetric Orbifolding and Non-Supersymmetric Heterotic theory}\label{sec:orbifold}

In this section, we review the supersymmetric heterotic string theory, and the procedure of orbifolding~\cite{Dixon:1985jw,Dixon:1986jc} to construct a non-supersymmetric theory from that. 
\subsection{Original Theories}

The Hilbert space for $(10-d)$-dimensional supersymmetric heterotic string theory with the toroidal compactification on the light-cone formalism is

\begin{equation}
\mathcal{H}=\hilb_{B}^{8-d,8-d}\otimes \hilb_{F}^{0,8}\otimes\hilb^{16+d,d}_{\mathrm{Internal}},
\end{equation}
where $\hilb_B^{8-d,8-d},\hilb_F^{0,8}$ are the bosonic or fermionic Fock spaces without GSO projection constructed by $\{\alpha_n^i,\tilde{\alpha}_n^i\}_{2\leq i\leq9-d}$ and $\{\tilde{\psi}_r^i\}_{2\leq i \leq 9}$, and  $\hilb^{16+d,d}_{\mathrm{Internal}}$ is a vector space constructed by an even self-dual lattice $\Gamma_{16+d,d}$ :
\begin{equation}
    \begin{aligned}
\hilb_{\mathrm{Internal}}^{16+d,d}=\hilb_B^{16+d,d}\otimes\qty(\bigoplus_{p\in\Gamma_{16+d,d}}\C\ket{p}).
    \end{aligned}
\end{equation}

For a point $p=(p_L,p_R)\in \Gamma_{16+d,d}$, the state $\ket{p}$ satisfies

\begin{equation}
    \begin{aligned}
        L_0\ket{p}=&\frac{1}{2}p_L^2\ket{p},\\
        \tilde{L}_0\ket{p}=&\frac{1}{2}p_R^2\ket{p}.
    \end{aligned}
\end{equation}
The torus partition function of a theory is 
\begin{equation}
\begin{aligned}
    Z^{(10-d)}_{\text{SUSY}}(\tau,\bar{\tau})
         =&Z_B^{(8-d)}\cdot \qty(\bar{V}_8-\bar{S}_8) \cdot  \frac{1}{\eta^{16+d}\bar{\eta}^d}\sum_{p\in\Gamma_{16+d,d}}q^{\frac{1}{2}p_L^2}\bar{q}^{\frac{1}{2}p_R^2},
\label{eq:SUSY_partition_function}\end{aligned}
\end{equation}
where $\tau\in \C$ with $\Im \tau>0$ is a complex modulus of a torus, and $q=\exp 2\pi i \tau$.  We have used  
\begin{equation}
\begin{aligned}
    Z_B^{(8-d)}:=&\frac{1}{(\Im \tau)^{\frac{8-d}{2}}}\tr_{\hilb_{B}^{8-d,8-d}}  q^{L_0-\frac{8-d}{24}}\bar{q}^{\tilde{L}_0-\frac{8-d}{24}}\\
   =&\frac{1}{(\Im \tau)^{\frac{8-d}{2}}}\frac{1}{(\eta\bar{\eta})^{8-d}},\\
\label{eq:ZB}\end{aligned}
\end{equation}
where $\eta$ is the Dedekind eta function. The factor $(\bar{V}_8-\bar{S}_8)$ comes from the worldsheet fermions, where $V_8, S_8$ are the $D_4$ character vector and spinor conjugacy classes, see App.~\ref{sec:theta} for the detail.

It follows from the Poisson summation formula that the modular invariance of the partition function requires the lattice $\Gamma_{16+d,d}$ to be even and self-dual:
\begin{equation}
\begin{aligned}
    Z^{(10-d)}_{\text{SUSY}}(\tau+1,\bar{\tau}+1)=&Z_B^{(8-d)}\cdot \qty(\bar{V}_8-\bar{S}_8) \cdot  \frac{1}{\eta^{16+d}\bar{\eta}^d}\sum_{p\in\Gamma_{16+d,d}}q^{\frac{1}{2}p_L^2}\bar{q}^{\frac{1}{2}p_R^2}e^{\pi i (p_L^2-p_R^2)},\\
    Z^{(10-d)}_{\text{SUSY}}\qty(-\frac{1}{\tau},-\frac{1}{\bar{\tau}})=&Z_B^{(8-d)}\cdot \qty(\bar{V}_8-\bar{S}_8) \cdot  \frac{1}{\eta^{16+d}\bar{\eta}^d}\sum_{p\in\Gamma_{16+d,d}^\ast}q^{\frac{1}{2}p_L^2}\bar{q}^{\frac{1}{2}p_R^2},
\end{aligned}
\end{equation}
where $\Gamma_{16+d,d}^\ast$ is the dual lattice of $\Gamma_{16+d,d}$. The modular T-invariance  of $Z^{(10-d)}_{\text{SUSY}}$ indicates that  $\Gamma_{16+d,d}$ is an even lattice: every $p=(p_L,p_R)\in\Gamma_{16+d,d}$ satisfies $p^2=p_L^2-p_R^2\in 2\Z$. The modular S-invariance of $Z^{(10-d)}_{\text{SUSY}}$ indicates that $\Gamma_{16+d,d}$ is a self-dual lattice.\footnote{Note that an even lattice $\Gamma_{16+d,d}$ satisfies $\Gamma_{16+d,d}\subset\Gamma_{16+d,d}^\ast$.} Since there are only the $E_8\times E_8$ root lattice and $\Spin(32)/\Z_2$ lattice for $(16,0)$ even self-dual lattices, it follows that there are only $E_8\times E_8$  and $\Spin(32)/\Z_2$ gauge symmetries in the 10-dimensional heterotic strings. 
In this way, even self-dual lattices play an important role in heterotic strings.

\subsection{Even Self-Dual Lattices and Gauge Symmetries}

What kinds of gauge group are possible in $(10-d)$-dimensional supersymmetric heterotic string theory? This can be understood by whether the root lattice of a gauge symmetry of rank $16+d$ can be embedded into an even self-dual lattice $\Gamma_{16+d,d}$.
Recently, all charge lattices corresponding to maximal gauge symmetry are identified for 9d and 8d cases~\cite{Font:2020rsk}.
In this subsection, we review the discussion of the paper.
 
Let $\g$ be a Lie algebra and $\Lambda_R(\g)$ be the root lattice of $\g$. A lattice $M$ is called an \textit{overlattice} of $\Lambda_R(\g)$ when $\Lambda_R(\g)\subset M\subset \Lambda_R(\g)^\ast$ and $\Q$-valued bilinear form of $\Lambda_R(\g)^\ast$ restricted on $M$ takes values in $\Z$.\footnote{Notice that $\Lambda_R(\g)^\ast$ is the dual lattice of $\Lambda_R(\g)$.}
Whether a Lie algebra is allowed in heterotic strings as its maximal gauge symmetry is determined by the following condition:
\begin{screen}
\textbf{Condition}

A root lattice $\Lambda_R(\g)$ has an embedding in an even self-dual lattice $\Pi_{d+16,d}$ if and only if $\Lambda_R(\g)$ has an overlattice $M$ with the following properties:

(1) there exists an even lattice $T$ of signature $(0,d)$ such that $(T^\ast/T,q_T)$ is isomorphic to $(M^\ast/M,q_M)$ up to $\Z$, where $q_T,q_M$ are the quadratic bilinear forms on $T^\ast/T,M^\ast/M$, respectively.

(2) The sublattice $M_{root}$ of $M$ coincides with $\Lambda_R(\g)$, where $M_{root}$ is the sublattice of $M$ generated by vectors of norm 2.
 
\end{screen}

All allowed groups of maximal rank are listed in Tables 11 and 12 of \cite{Font:2020rsk}.

How to construct an even self-dual lattice from $M$ and $T$, satisfying this condition? Let $x^{(M)}_i\in M^\ast,x^{(T)}_i\in T^\ast, i=1,\cdots,d$ be the generators of $M^\ast/M, T^\ast/T$, respectively, and $q_M\qty(x_i^{(M)})-q_T\qty(x_i^{(T)})\in 2\Z$. Then, the following $(16+d,d)$ lattice $\Gamma_{16+d,d}$ is self-dual:
\begin{equation}
    \Gamma_{16+d,d}=M\oplus T +\Z\qty(x_1^{(M)};x_1^{(T)})+\cdots + \Z\qty(x_d^{(M)};x_d^{(T)}),
\end{equation}
where $\Z\qty(x_i^{(M)};x_i^{(T)})$ means the one-dimensional lattice generated by $\qty(x_i^{(M)};x_i^{(T)}) \in M^\ast\oplus T^\ast$. Here $\oplus$ means a disjoint union of two abelian groups, and $+$ means the sum of two abelian subgroups of $M^\ast\oplus T^\ast$.

\subsection{Asymmetric Orbifolding}
To discuss orbifolding, suppose that $\Gamma_{16+d,d}$ possesses a $\Z_2$ symmetry $g$. The following $(8+d,d)$ lattice $I_{8+d,d}$ constructed by the action of $g$ on $\Gamma_{16+d,d}$ is called \textit{Invariant Lattice}:
\begin{equation}
    I_{8+d,d}\coloneqq\{ x\in\Gamma_{16+d,d} | g(x)=x\}.
\end{equation}
For example, the lattice $\Gamma_{16,0}=\Lambda_R(E_8)\oplus \Lambda_R(E_8)$ has the symmetry $g:(x_1,x_2)\mapsto (x_2,x_1)$, where $\Lambda_R(E_8)$ is the root lattice of $E_8$. Then the invariant lattice is given as follows:
\begin{equation}
    \begin{aligned}
        I_{8+d,d}=&\{(x,x)| x\in\Lambda_R(E_8)\}\\
    \cong &\sqrt{2}\Lambda_R(E_8),
    \end{aligned}
\end{equation}
where $\cong $ means an isomorphism of the abelian group. This leads to the construction of a new theory based on $I_{8+d,d}$, as we will see in the following subsections. This process is called \textit{Asymmetric Orbifolding}~\cite{Narain:1986qm}, since $g$ acts only left part of the lattice.

A new theory consists of two sectors: the untwisted sector and the twisted sector. The untwisted sector is what remains after the projection and is part of the original theory. In contrast, the twisted sector emerges in the course of asymmetric orbifolding, and its existence is required by the modular invariance, so it is added.
The untwisted  is based on a lattice $I_{8+d,d}\subset \Gamma_{16+d,d}$, called \textit{invariant lattice} ,the latter appears along with the dual lattice $I^\ast_{8+d,d}$ for the modular invariance of the partition function.
\subsection{Untwisted Sector}

$\Z_2$ symmetry $g$ of lattice acts a state $\ket{x_1,x_2,x}\otimes\ket{p}\in\qty(H^{8,0}_{B}\otimes H^{8,0}_{B}\otimes H^{d,d}_{B})\otimes \hilb_{\Gamma_{16+d,d}}=\hilb^{16+d,d}_{\mathrm{Internal}}$ as follows:

\begin{equation}
    g\qty(\ket{x_1,x_2,x}\otimes \ket{p})=\ket{x_2,x_1,x}\otimes\ket{g(p)}.
\end{equation}

With this action of $g$, Hilbert space of the untwisted sector is given by the projection:
\begin{equation}
\mathcal{H}^{(\text{untwisted})}=\frac{1+g(-1)^F}{2}\mathcal{H}
\end{equation}
where $F$ is the spacetime fermion number (not worldsheet), which acts on $\mathcal{H}_{\mathrm{Fermion}}$.

The trace over $\hilb^{16+d,d}_{\mathrm{Internal}}$ with the insertion of $g$ can be computed as follows:

\begin{equation}
    \begin{aligned}
        &\tr_{\hilb^{16+d,d}_{\mathrm{Internal}}} g q^{L_0-\frac{16+d}{24}}\bar{q}^{\tilde{L}_0-\frac{16+d}{24}}\\
        =&\tr_{\hilb^{16+d,d}_B} g q^{L_0-\frac{16+d}{24}}\bar{q}^{\tilde{L}_0-\frac{16+d}{24}} \cdot \tr_{\hilb^{16+d,d}_{\Gamma_{16+d,d}}} g q^{L_0}\bar{q}^{\tilde{L}_0}\\
        =&\frac{1}{\eta(q)^{8+d}\bar{\eta}(\bar{q})^d \eta(q^2)^8} \cdot \sum_{(p_L,p_R)\in I_{16+d,d}} q^{\frac{1}{2}p_L^2}\bar{q}^{\frac{1}{2}p_R^2}\\
        =&\frac{1}{\eta^{16+d}\bar{\eta}^d}\qty(\frac{\eta^3}{\theta_2})^4\sum_{(p_L,p_R)\in I_{16+d,d}} q^{\frac{1}{2}p_L^2}\bar{q}^{\frac{1}{2}p_R^2}
        \end{aligned}
\end{equation}
Therefore, the partition function of the untwisted sector is 
\begin{equation}
    Z^{\mathrm{(untwisted)}}=\frac{1}{2}Z_B^{(8-d)}\qty\big((\bar{V}_8-\bar{S}_8)Z(1,1)+(\bar{V}_8+\bar{S}_8)Z(1,g)),
\end{equation}
where
\begin{equation}
    \begin{aligned}
         Z(1,1)\coloneqq&\tr_{\hilb^{16+d,d}_{\mathrm{Internal}}} q^{L_0-\frac{16+d}{24}}\bar{q}^{\tilde{L}_0-\frac{d}{24}}
         =\frac{1}{\eta^{16+d}\bar{\eta}^{d}}\sum_{p\in \Gamma_{16+d,d}}q^{\frac{1}{2}p_R^2}\bar{q}^{\frac{1}{2}p_R^2},\\
Z(1,g)\coloneqq&\tr_{\hilb^{16+d,d}_{\mathrm{Internal}}} g q^{L_0-\frac{16+d}{24}}\bar{q}^{\tilde{L}_0-\frac{d}{24}}
         =\frac{1}{\eta^{16+d}\bar{\eta}^d}\qty(\frac{\eta^3}{\theta_2})^4\sum_{p\in I_{8+d,d}}q^{\frac{1}{2}p_R^2}\bar{q}^{\frac{1}{2}p_R^2}.
    \end{aligned}
\end{equation}
The sign of $S_8$ changes between the first and second terms due to the effect of $(-1)^F$.

It can be seen that the dilatino and the gravitino are projected out, therefore there is no supersymmetry in this new theory.

\subsection{Twisted Sector}

The partition function of string theory must be modular invariant but 
$Z^{\mathrm{(untwisted)}}$ is not, because $I_{8+d,d}$ is not an even self-dual lattice (it is even but not self-dual). In fact, $Z(1,g)$ is not invariant under $\tau\to -1/\tau$. Therefore we must add appropriate terms to $Z^{\mathrm{(untwisted)}}$ in order to obtain a modular invariant partition function. The spectrum can be read off of it. 

The untwisted sector partition function is invariant under modular $T$ transformation because of
\begin{equation}
\qty(\bar{V}_8+\bar{S}_8)(\bar{\tau}+1)Z(1,g)\qty(\tau+1,\bar{\tau}+1)=\qty(\bar{V}_8+\bar{S}_8)(\bar{\tau})Z(1,g)\qty(\tau,\bar{\tau}),
\end{equation}
where \eqref{eq:fermions_modular_tr}\eqref{eq:eta_modular_tr} have been used.
We define $Z(g,1)(\tau,\bar{\tau}), Z(g,g)(\tau,\bar{\tau})$ by the modular transformation as follows:\footnote{There is no discrete torsion as $H_2(\Z_2,\U(1))=0$~\cite{Vafa:1986wx}.}
\begin{equation}
\begin{aligned}
    (\bar{O}_8-\bar{C}_8)Z(g,1)\coloneqq&\qty(\bar{V}_8+\bar{S}_8)\qty(-1/\bar{\tau})Z(1,g)\qty(-1/\tau,-1/\bar{\tau}),\\
    -\qty((\bar{O}_8+\bar{C}_8)Z(g,g))\coloneqq&(\bar{O}_8-\bar{C}_8)\qty(\bar{\tau}+1)Z(g,1)\qty(\tau+1,\bar{\tau}+1).
\end{aligned}
\end{equation}
Here, $\bar{O}_8$ and $\bar{C}_8$ are defined in \eqref{eq:D4_characters}.
From the Poisson summation formula, we obtain
\begin{equation}
    \begin{aligned}
        Z(g,1)=&\frac{1}{\eta^{16+d}\bar{\eta}^d} \qty(\frac{\eta^3}{\theta_4})^4 \sum_{p\in I_{8+d,d}^\ast}q^{\frac{1}{2}p_L^2}\bar{q}^{\frac{1}{2}p_R^2},\\
        Z(g,g)=&\frac{1}{\eta^{16+d}\bar{\eta}^d} \qty(\frac{\eta^3}{\theta_3})^4 \sum_{p\in I_{8+d,d}^\ast}q^{\frac{1}{2}p_L^2}\bar{q}^{\frac{1}{2}p_R^2}e^{\pi i p^2},\\
    \end{aligned}
\end{equation}
where $p^2=p_L^2-p_R^2$. Here we used the modular transformation property of theta functions, see App.~\ref{sec:theta}.

By adding all of these together, the partition function of the 9d non-supersymmetric theory constructed by this orbifolding is given as follows:
\begin{equation}
\begin{aligned}
        Z^{(10-d)}_{\cancel{\text{SUSY}}}=&\frac{1}{2}Z^{(8-d)}_B\left((\bar{V}_8-\bar{S}_8)Z(1,1)+(\bar{V}_8+\bar{S}_8)Z(1,g)\right.\\
        +& \left.(\bar{O}_8-\bar{C}_8)Z(g,1)-(\bar{O}_8+\bar{C}_8)Z(g,g)\right).
\end{aligned}
\end{equation}

\subsection{Modular Invariance of the Partition Function}
The total partition function is given as follows:
\begin{equation}
\begin{aligned}
Z_{\cancel{\text{SUSY}}}^{(10-d)} = \frac{1}{2}\frac{1}{(\Im \tau)^{\frac{8-d}{2}}}  \frac{1}{\bar{\eta}^8\eta^{24}} \Bigg\{ &\bar{V}_8 \left( \sum_{p\in \Gamma_{16+d,d}}  q^{\frac{1}{2}p_L^2}\bar{q}^{\frac{1}{2}p_R^2} 
+ \left( \frac{2 \eta^3}{\theta_2} \right)^4 \sum_{p\in I_{8+d,d}} q^{\frac{1}{2}p_L^2}\bar{q}^{\frac{1}{2}p_R^2}  \right) \\
- &\bar{S}_8 \left( \sum_{p\in \Gamma_{16+d,d}} q^{\frac{1}{2}p_L^2}\bar{q}^{\frac{1}{2}p_R^2} 
- \left( \frac{2 \eta^3}{\theta_2} \right)^4 \sum_{p\in I_{8+d,d}} q^{\frac{1}{2}p_L^2}\bar{q}^{\frac{1}{2}p_R^2}  \right) \\
+& \bar{O}_8 \sum_{p\in I_{8+d,d}^\ast} q^{\frac{1}{2}p_L^2}\bar{q}^{\frac{1}{2}p_R^2} 
\left( \left( \frac{\eta^3}{\theta_4} \right)^4 + \left( \frac{\eta^3}{\theta_3} \right)^4 e^{\pi i p^2} \right) \\
 - &\bar{C}_8 \sum_{p\in I_{8+d,d}^\ast} q^{\frac{1}{2}p_L^2}\bar{q}^{\frac{1}{2}p_R^2} 
\left( \left( \frac{\eta^3}{\theta_4} \right)^4 - \left( \frac{\eta^3}{\theta_3} \right)^4 e^{\pi i p^2} \right) 
\Bigg\}.
\end{aligned}
\end{equation}
The modular invariance of $Z^{(10-d)}_{\cancel{\text{SUSY}}}$ can be easily shown. It is enough to show that
\begin{equation}
    -(\bar{O}_8+\bar{C}_8)\qty(-\frac{1}{\bar{\tau}})Z(g,g)\qty(-\frac{1}{\tau},-\frac{1}{\bar{\tau}})=-(\bar{O}_8+\bar{C}_8)(\bar{\tau})Z(g,g)\qty(\tau,\bar{\tau}).
\end{equation}
It can be seen by the invariance of $(\bar{V}_8+\bar{S}_8)Z(1,g)$ and $(\bar{O}_8-\bar{C}_8)Z(g,1)$ under $\tau\to\tau+1$ and $\tau\to\tau+2$, respectively.
This can be thought of as the following condition:
\begin{equation}
I_{8+d,d}^\ast\bigr|_{\mathrm{even}}=I_{8+d,d},
\end{equation}
where 
\begin{equation}
    I_{8+d,d}^\ast\bigr|_{\mathrm{even}}:=\{ (p_L,p_R)\in I_{8+d,d}^\ast| p_L^2-p_R^2\in 2\Z\}.
\end{equation}

\subsection{Spectrum}
The twisted sector part of the partition function is as follows:
\begin{equation}
    \begin{aligned}
        Z^{\mathrm{(twisted)}}=&\frac{1}{2}Z_B^{(8-d)}\qty((\bar{O}_8-\bar{C}_8)Z(g,1)-(\bar{O}_8+\bar{C}_8)Z(g,g))\\
        =& \bar{O}_8 \sum_{p\in I_{8+d,d}^\ast} q^{\frac{1}{2}p_L^2}\bar{q}^{\frac{1}{2}p_R^2} 
\left( \left( \frac{\eta^3}{\theta_4} \right)^4 + \left( \frac{\eta^3}{\theta_3} \right)^4 e^{\pi i p^2} \right) \\
 - &\bar{C}_8 \sum_{p\in I_{8+d,d}^\ast} q^{\frac{1}{2}p_L^2}\bar{q}^{\frac{1}{2}p_R^2} 
\left( \left( \frac{\eta^3}{\theta_4} \right)^4 - \left( \frac{\eta^3}{\theta_3} \right)^4 e^{\pi i p^2} \right). 
    \end{aligned}
\end{equation}
The spectrum of the twisted sector can be read off from this partition function. Due to the level matching condition, we only need to consider terms where the powers of $q$ and $\bar{q}$ are matched. Among the lighter states, $(q\bar{q})^{-\frac{1}{2}}$ represents tachyons, and $(q\bar{q})^0$ represents massless particles.

Tachyonic states in the scalar conjugacy class can be read off as follows:
\begin{equation}
    \begin{aligned}
        &\frac{1}{2}\bar{O}_8 \sum_{p\in I_{8+d,d}^\ast} q^{\frac{1}{2}p_L^2}\bar{q}^{\frac{1}{2}p_R^2} 
\qty( \qty( \frac{\eta^3}{\theta_4} )^4 + \qty( \frac{\eta^3}{\theta_3} )^4 e^{\pi i p^2} )\\
\sim&\frac{1}{2}\bar{O}_8 \sum_{p\in I_{8+d,d}^\ast}q^{\frac{1}{2}p_L^2}\bar{q}^{\frac{1}{2}p_R^2}q^{\frac{1}{2}}\qty(\qty(1+8q^{\frac{1}{2}})+\qty(1-8q^{\frac{1}{2}})e^{\pi ip^2})\\
\sim&q^\frac{1}{2}\bar{O}_8. 
    \end{aligned}
\end{equation}
Massless states in conjugate spinor conjugacy class can be read off as follows
\begin{equation}
    \begin{aligned}
        &\frac{1}{2}\bar{C}_8 \sum_{p\in I_{8+d,d}^\ast} q^{\frac{1}{2}p_L^2}\bar{q}^{\frac{1}{2}p_R^2} 
\qty( \qty( \frac{\eta^3}{\theta_4} )^4 - \qty( \frac{\eta^3}{\theta_3} )^4 e^{\pi i p^2} )\\
\sim&\frac{1}{2}\bar{C}_8 \sum_{p\in I_{8+d,d}^\ast}q^{\frac{1}{2}p_L^2}\bar{q}^{\frac{1}{2}p_R^2}q^{\frac{1}{2}}\qty(\qty(1+8q^{\frac{1}{2}})-\qty(1-8q^{\frac{1}{2}})e^{\pi ip^2})\\
\sim &q^1\qty(8+n_{(1,0)})\bar{C}_8,
    \end{aligned}
\end{equation}
where $n_{(1,0)}$ is the number of elements $(p_L,p_R)\in I_{8+d,d}^\ast$ which satisfy $p_L^2=1, p_R^2=0$,
and we have used the following expansions:
\begin{equation}
\begin{aligned}
    \qty(\frac{2\eta^3}{\theta_2})^4=&1-16q+112q^2+O(q^3),\\
    \qty(\frac{\eta^3}{\theta_3})^4=&q^{\frac{1}{2}}\qty(1-8q^{\frac{1}{2}})+q^{\frac{1}{2}}O(q),\\
    \qty(\frac{\eta^3}{\theta_4})^4=&q^{\frac{1}{2}}\qty(1+8q^{\frac{1}{2}})+q^{\frac{1}{2}}O(q),\\
\end{aligned}
\end{equation}
and
\begin{equation}
    \begin{aligned}
        &\bar{O}_8=\frac{1}{\bar{\eta}^4}\qty(1+24\bar{q}+\cdots),
        &&\bar{C}_8=\frac{1}{\bar{\eta}^4}\qty(8\bar{q}^{\frac{1}{2}}+\cdots).
    \end{aligned}
\end{equation}
Note that $\eta$ is given in \eqref{eq:eta_function}. For these materials, see App.~\ref{sec:theta} for details.

\subsection{Folding of Dynkin Diagram}
As an example, we describe the folding $A_{2n-1}\to C_n$. Let us consider the root lattice of $A_{2n-1}$:
\begin{equation}
    \Lambda_R(A_{2n-1})=\bigoplus_{i=1}^{2n-1}\Z\alpha_i^{(A_{2n-1})}.
\end{equation}
The specific form of $\alpha_i^{(A_{2n-1})}$ is summarized in the App.~\ref{sec:lie algebra}. This lattice possesses the following symmetry:
\begin{equation}
\begin{aligned}
        g:&\sum_{i=1}^{2n-1}n_i \alpha_i^{(A_{2n-1})}
        \mapsto  \sum_{i=1}^{2n-1}n_{2n-i} \alpha_i^{(A_{2n-1})}.
\end{aligned}
\end{equation}

The new lattice obtained by taking the $g$-invariant part of $\Lambda_R(A_{2n-1})$ is close to the root lattice of $C_n$:
\begin{equation}
\begin{aligned}
        \Lambda_R^g(A_{2n-1}):=&\{ x\in \Lambda_R(A_{2n-1})| g(x)=x\}\\
        =&\sqrt{2}\Z\alpha_1^{(C_{n})}\oplus\cdots\oplus\sqrt{2}\Z\alpha_{n-1}^{(C_n)}\oplus\Z\frac{1}{\sqrt{2}}\alpha_n^{(C_n)}\\
        \cong&\sqrt{2}\qty(\Lambda_R(C_n)+\frac{1}{2}\Z\alpha_n^{(C_n)}).
\end{aligned}
\end{equation}
This process is called \textit{folding}, since it corresponds to the folding of the Dynkin diagram:

\begin{figure}[h]
    \centering
    \includegraphics[width=0.4\linewidth]{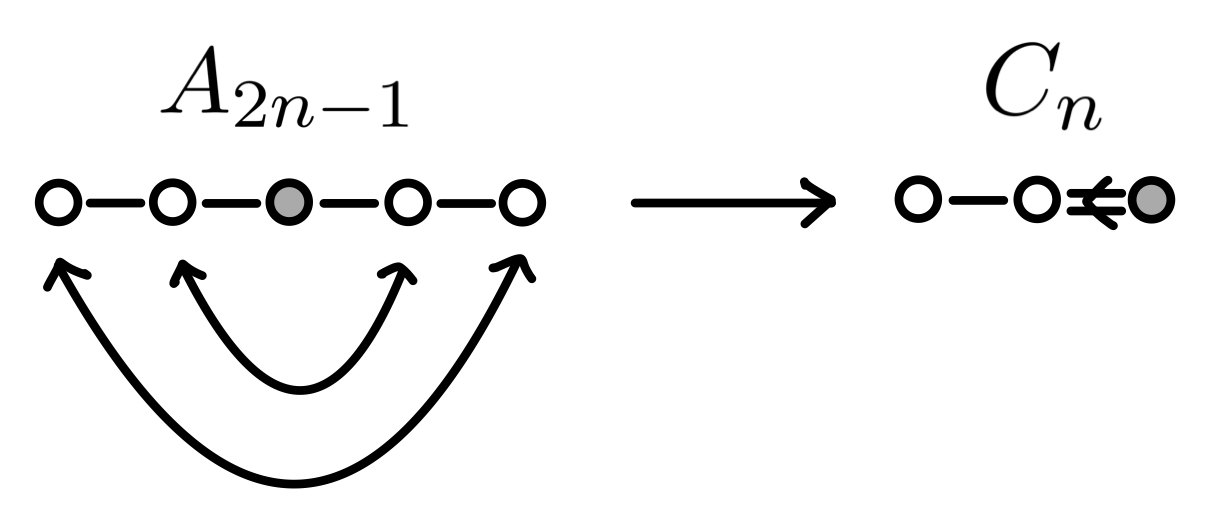}
    \caption{Folding from $A_{2n-1}$ to $C_n$}
    \label{fig:enter-label}
\end{figure}

Root basis and fundamental weights of $A_{2n-1}$ and $C_n$ have following relationships:
\begin{equation}
    \begin{aligned}
        &\alpha_i^{(C_n)}=\frac{1}{\sqrt{2}}\qty(\alpha_i^{(A_{2n-1})}+\alpha_{2n-i}^{(A_{2n-1})}),\quad\text{for}\quad1\leq i\leq n-1,\\
    &\alpha_n^{(C_n)}=\sqrt{2}\alpha_n^{(A_{2n-1})},\\
    &\omega_i^{(C_n)}=\frac{1}{\sqrt{2}}\qty(\omega_i^{(A_{2n-1})}+\omega_{2n-i}^{(A_{2n-1})}),\quad\text{for}\quad 1\leq i\leq n-1,\\
    &\omega_n^{(C_n)}=\sqrt{2}\omega_n^{(A_{2n-1})}.
    \end{aligned}
\end{equation}

Besides $A_{2n-1}\to C_n$, there are several other types of  folding of Dynkin diagrams:

\begin{equation}
    \begin{aligned}
        &A_{2n}\to B_{n},
        &&D_{n} \to B_{n-1},
        &&D_4   \to G_2,
        &&E_6   \to F_4,
    \end{aligned}
\end{equation}
but none of them appears in this paper.

\section{Non-Supersymmetric Strings from Supersymmetric Strings}\label{sec:fermionization}

In this section, we argue the relation between non-supersymmetric and supersymmetric heterotic strings in 9d, based on \cite{Karch:2019lnn,Tachikawa:2018}.
Suppose that we have a 2d theory $T$ with a $0$-form $G=\Z_2$ symmetry, and we orbifold the $T$ theory by $G$. The torus partition function of the orbifolded theory $\hat{T}=T/\Z_2$ is
\begin{align}
    Z_{\hat{T}}(1,1)&=\frac{1}{2}\left(Z_{T}(1,1)+Z_{T}(1,g)+Z_{T}(g,1)+Z_{T}(g,g)\right),
\label{eq:Z_That(1,1)}\end{align}
where $Z_{T}(g,h)$ is the partition function of $T$ with $g$ and $h$ twists along worldsheet space and time directions, respectively.

It is known that when the $\hat{T}=T/\Z_2$  theory has a dual $0$-form $\hat{G}=\Z_2$ symmetry that acts non-trivially on the twisted sector states~\cite{Vafa:1989ih}. Therefore, $\hat{T}$ torus partition function with $\hat{g}$ twist along the worldsheet time direction is 
\begin{align}
    Z_{\hat{T}}(1,\hat{g})&=\frac{1}{2}\left(Z_{T}(1,1)+Z_{T}(1,g)-Z_{T}(g,1)-Z_{T}(g,g)\right).
\label{eq:Z_That(1,g)}\end{align}
The partition function of other boundary conditions can be derived from the modular transformations:
\begin{align}
    Z_{\hat{T}}(\hat{g},1)&=\frac{1}{2}\left(Z_{T}(1,1)-Z_{T}(1,g)+Z_{T}(g,1)-Z_{T}(g,g)\right),
    \nonumber\\
    Z_{\hat{T}}(\hat{g},\hat{g})&=\frac{1}{2}\left(Z_{T}(1,1)-Z_{T}(1,g)-Z_{T}(g,1)+Z_{T}(g,g)\right).
\label{eq:Z_That_twisted}\end{align}
From Eqs.~\eqref{eq:Z_That(1,1)}\eqref{eq:Z_That(1,g)}\eqref{eq:Z_That_twisted}, we observe that the $\mathbb{Z}_2$ orbifolding of $\hat{T}$ theory turns out to be the original $T$ theory, that is $(T/G)/\hat{G}=T$:
\begin{align}
    Z_{T}(1,1)=\frac{1}{2}\left(Z_{\hat{T}}(1,1)+Z_{\hat{T}}(1,\hat{g})+Z_{\hat{T}}(\hat{g},1)+Z_{\hat{T}}(\hat{g},\hat{g})\right).    
\end{align}
Summarizing so far, we have learned 
\begin{align}
    \hat{T}=T/G\quad\text{and}\quad T=\hat{T}/\hat{G}.
\label{eq:equivalence}\end{align}

In our case, we are interested in a non-supersymmetric theory whose torus partition function is
\begin{align}
    Z_{{\cancel{\text{SUSY}}}}=\frac{Z_B^{(8-d)}}{\eta^{16+d}\bar{\eta}^d} \left(c_{\bar{O}_8}(\tau) \bar{O}_8 + c_{\bar{V}_8}(\tau) \bar{V}_8 + c_{\bar{S}_8}(\tau) \bar{S}_8 + c_{\bar{C}_8}(\tau) \bar{C}_8\right).
\label{eq:non-SUSY_partition_function}\end{align}
We can massage the expression as
\begin{align}
    Z_{{\cancel{\text{SUSY}}}}=&Z_B^{(8-d)} \left(\frac{1}{2}(\bar{V}_8-\bar{S}_8)(c_{\bar{V}_8}(\tau)-c_{\bar{S}_8}(\tau))+\frac{1}{2}(\bar{V}_8+\bar{S}_8)(c_{\bar{V}_8}(\tau)+c_{\bar{S}_8}(\tau))\right.
    \nonumber\\
    &\left.+\frac{1}{2}(\bar{O}_8-\bar{C}_8)(c_{\bar{O}_8}(\tau)-c_{\bar{C}_8}(\tau))+\frac{1}{2}(\bar{O}_8+\bar{C}_8)(c_{\bar{O}_8}(\tau)+c_{\bar{C}_8}(\tau))\right).
\end{align}
With the modular transformation law of $D_4$ characters~\eqref{eq:fermions_modular_tr} and $Z_B^{(8-d)}$~\eqref{eq:ZB}\eqref{eq:eta_modular_tr}, the modular S-invariance requires
\begin{align}
    &\left.c_{\bar{V}_8}-c_{\bar{S}_8}\right|_{-\tau^{-1}}=\left.c_{\bar{V}_8}-c_{\bar{S}_8}\right|_{\tau}\times\frac{1}{\tau^8(|\tau|^2)^{d/2}},
    \nonumber\\
    &\left.c_{\bar{V}_8}+c_{\bar{S}_8}\right|_{-\tau^{-1}}=\left.c_{\bar{O}_8}-c_{\bar{C}_8}\right|_{\tau}\times\frac{1}{\tau^8(|\tau|^2)^{d/2}},
    \nonumber\\
    &\left.c_{\bar{O}_8}+c_{\bar{C}_8}\right|_{-\tau^{-1}}=\left.c_{\bar{O}_8}+c_{\bar{C}_8}\right|_{\tau}\times\frac{1}{\tau^8(|\tau|^2)^{d/2}}.
\end{align}
Similarly, from the modular T-invariance, we get
\begin{align}
    &\left.(\bar{O}_8-\bar{C}_8)(c_{\bar{O}_8}-c_{\bar{C}_8})\right|_{\tau+1}
    =\left.(\bar{O}_8+\bar{C}_8)(c_{\bar{O}_8}+c_{\bar{C}_8})\right|_{\tau},
    \nonumber\\
    &\left.(\bar{O}_8+\bar{C}_8)(c_{\bar{O}_8}+c_{\bar{C}_8})\right|_{\tau+1}
    =\left.(\bar{O}_8-\bar{C}_8)(c_{\bar{O}_8}-c_{\bar{C}_8})\right|_{\tau}.
\end{align}

We orbifold the theory by $(-1)^{f}$ which nontrivially acts on $\bar{O}_8$ and $\bar{C}_8$. The partition function of orbifolded theory $\hat{T}=\cancel{\text{SUSY}}/(-1)^{f}$ is
\begin{align}
    Z_{\hat{T}}(1,1)=&Z_B^{(8-d)} \left(\frac{1}{2}(\bar{V}_8-\bar{S}_8)(c_{\bar{V}_8}(\tau)-c_{\bar{S}_8}(\tau))+\frac{1}{2}(\bar{V}_8+\bar{S}_8)(c_{\bar{V}_8}(\tau)+c_{\bar{S}_8}(\tau))\right.
    \nonumber\\
    &\left.+\frac{1}{2}(\bar{O}_8-\bar{C}_8)(c_{\bar{O}_8}(\tau)-c_{\bar{C}_8}(\tau))+\frac{1}{2}(\bar{O}_8+\bar{C}_8)(c_{\bar{O}_8}(\tau)+c_{\bar{C}_8}(\tau))\right),
    \nonumber\\
    Z_{\hat{T}}(1,(-1)^{f})=&Z_B^{(8-d)} \left(\frac{1}{2}(\bar{V}_8-\bar{S}_8)(c_{\bar{V}_8}(\tau)-c_{\bar{S}_8}(\tau))+\frac{1}{2}(\bar{V}_8+\bar{S}_8)(c_{\bar{V}_8}(\tau)+c_{\bar{S}_8}(\tau))\right.
    \nonumber\\
    &\left.-\frac{1}{2}(\bar{O}_8-\bar{C}_8)(c_{\bar{O}_8}(\tau)-c_{\bar{C}_8}(\tau))
    -\frac{1}{2}(\bar{O}_8+\bar{C}_8)(c_{\bar{O}_8}(\tau)+c_{\bar{C}_8}(\tau))\right),
    \nonumber\\
    Z_{\hat{T}}((-1)^{f},1)=&Z_B^{(8-d)} \left(\frac{1}{2}(\bar{V}_8-\bar{S}_8)(c_{\bar{V}_8}(\tau)-c_{\bar{S}_8}(\tau))-\frac{1}{2}(\bar{V}_8+\bar{S}_8)(c_{\bar{V}_8}(\tau)+c_{\bar{S}_8}(\tau))\right.
    \nonumber\\
    &\left.+\frac{1}{2}(\bar{O}_8-\bar{C}_8)(c_{\bar{O}_8}(\tau)-c_{\bar{C}_8}(\tau))
    -\frac{1}{2}(\bar{O}_8+\bar{C}_8)(c_{\bar{O}_8}(\tau)+c_{\bar{C}_8}(\tau))\right),
    \nonumber\\
    Z_{\hat{T}}((-1)^{f},(-1)^{f})=&Z_B^{(8-d)} \left(\frac{1}{2}(\bar{V}_8-\bar{S}_8)(c_{\bar{V}_8}(\tau)-c_{\bar{S}_8}(\tau))
    -\frac{1}{2}(\bar{V}_8+\bar{S}_8)(c_{\bar{V}_8}(\tau)+c_{\bar{S}_8}(\tau))\right.
    \nonumber\\
    &\left.-\frac{1}{2}(\bar{O}_8-\bar{C}_8)(c_{\bar{O}_8}(\tau)-c_{\bar{C}_8}(\tau))
    +\frac{1}{2}(\bar{O}_8+\bar{C}_8)(c_{\bar{O}_8}(\tau)+c_{\bar{C}_8}(\tau))\right).
\end{align}
Consequently, the partition function of the orbifolded theory is
\begin{align}
    Z_{\hat{T}}=Z_B^{(8-d)} \left(\frac{1}{2}(\bar{V}_8-\bar{S}_8)(c_{\bar{V}_8}(\tau)-c_{\bar{S}_8}(\tau))\right).
\end{align}
This partition function describes a supersymmetric theory~\eqref{eq:SUSY_partition_function} since the presence of the graviton leads to the presence of the gravitino. 
Therefore, we have seen 
\begin{align}
    [\cancel{\text{SUSY}}\text{ theory}]/(-1)^{f}=\text{SUSY theory}.
\end{align}
Combined with the general statement~\eqref{eq:equivalence}, we observe that the non-supersymmetric heterotic theory is obtained by gauging $\Z_2$ of the supersymmetric heterotic theory as long as the form of the torus partition function of the non-supersymmetric theory is ~\eqref{eq:non-SUSY_partition_function}.

In this paper, we consider the non-supersymmetric heterotic string theory obtained from $S^1$ compactification of the 10d $E_8$ string (see Fig.~\ref{fig:theories}), known as $B_{III}$ theory. We also consider several examples for 8d cases in Sec.~\ref{sec:8d}. Since the form of the partition function is Eq.~\eqref{eq:non-SUSY_partition_function}, we can identify all gauge group and matter fields of the theory from the $\Z_2$ orbifold of the 9d supersymmetric theory.
We will perform this program in Sec.~\ref{sec:9d}. Furthermore, we can learn about the disconnected part of the spacetime gauge group through the identification of symmetry from which we can predict various branes. 

\section{Disconnected Gauge Group and Non-Supersymmetric Branes}\label{sec:branes}
In Sec.~\ref{sec:fermionization}, we have learned that the non-supersymmetric theory is obtained by $\Z_2$ gauging of the supersymmetric theory.
Here, we argue that this $\Z_2$ symmetry can also be used to predict new objects in supersymmetric theory.

\begin{figure}[h]
    \centering
    \begin{tikzpicture}

        \draw[thick] (6.5-3,-0.5+0.3) to[out=90+15, in=-90+15] (6.5-3,3+0.3); 
        \node at (6.2-3,2.0+0.3) {$\mathbf{N}$}; 
        \node at (6.9-3,2.05+0.3) {$\overline{\mathbf{N}}$}; 
        \filldraw[blue] (6.2-3,1.5+0.3) circle (3pt); 
        \filldraw[blue] (6.8-3,1.5+0.3) circle (3pt); 
        \filldraw[red] (3.5,3.3) circle (3pt); 
        \node[red] at (3.5,3.7) {$(D-3)$-brane}; 
        
        \draw[->, blue, thick] (6.2-3, 1.3+0.3) arc (-235:45:0.5);

    \end{tikzpicture}
    \caption{A codimension two Gukov-Witten operator (black line) charged under a dual $(D-2)$-form symmetry generates a holonomy along the transverse $S^1$ direction. The Gukov-Witten operator can be viewed as an insertion of the probe vortex with the holonomy. Consequently, this operator can end at the dynamical $(D-3)$-brane(Alice string/twisted vortex) represented by the red dot. 
    The charge of the vortex is measured by the homotopy group $\pi_0$ or the bordism group $\Omega_1$.}
    \label{fig:brane_diagram}
\end{figure}
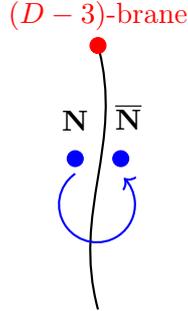

In the next section, we will see that the charge lattice admits outer automorphism $\Z_2$ symmetry for several cases. In light of no global symmetry conjecture (or more refined cobordism conjecture~\cite{McNamara:2019rup}), we should view this $\Z_2$ symmetry as the gauge symmetry in the bulk. Consequently, the bulk gauge symmetry becomes disconnected such as $\SU(N)\rtimes\Z_2$, called principal extensions in the literature~\cite{Siebenthal1956/57,wendt1999weylscharacterformulanonconnected,Bachas:2000ik,Maldacena:2001xj,Stanciu:2001vw}. That is, we start from $\SU(N)$ gauge theory, and then gauge the $0$-form charge conjugation symmetry. In general, gauging the $p$-form symmetry in $D$-dimensional theory leads to a theory with a dual $(D-p-2)$-form symmetry~\cite{Gaiotto:2014kfa,Tachikawa:2017gyf}. In the case at hand, we have the dual $(D-2)$-form symmetry. The charged object is a codimension two Gukov-Witten operator~\cite{Gukov:2006jk,Gukov:2008sn} (see Fig.~\ref{fig:brane_diagram}) while the symmetry operator is a topological Wilson line. Given no global symmetry conjecture, we would like to explicitly break $(D-2)$-form symmetry, which is realized by $(D-3)$-branes. The $(D-3)$-branes that break the dual $(D-2)$-form symmetry are viewed as a generalization of the Alice string~\cite{Schwarz:1982ec} in $\mathrm{O}(2)$ gauge theory (or twisted vortex in \cite{Heidenreich:2021xpr}). Recently, this is used to predict new non-supersymmetric branes in type IIB~\cite{Dierigl:2022reg} and heterotic string~\cite{Kaidi:2023tqo}(see also \cite{Alvarez-Garcia:2024vnr}) for $D=10$.\footnote{A $(-1)$-brane in the heterotic string is proposed in \cite{Alvarez-Garcia:2024vnr}.} The type IIB R$7$-brane is related to $\Z_2$ symmetry known as $\Omega$ or $(-1)^{F_L}$, and heterotic $7$-brane is related to exchange of two $E_8$'s. Our case corresponds to $D=9$ and $8$.

Similarly, the first homotopy group of the gauge symmetry is related to new branes. We will see that if the spacetime gauge symmetry is $\SU(N)\rtimes\Z_2$, the spectrum is not complete. In this case, there exists a non-invertible $1$-form symmetry~\cite{Arias-Tamargo:2022nlf,Bhardwaj:2022yxj}. Since we believe that all the states electrically charged under the gauge group are constructed as states on the fundamental heterotic string, we should view the gauge symmetry as $\SU(N)\rtimes\Z_2/\Z_k$ so that the electric charge completeness is achieved. Next, we consider the magnetic charge completeness. 
Without the charge conjugation gauge symmetry, 't Hooft operator is classified by $\pi_1(\SU(N)\rtimes\Z_2/\Z_k)=\Z_k$. This is nothing but magnetic $(D-3)$-from symmetry, and the magnetic $(D-4)$-brane is predicted in order to break it~\cite{Kaidi:2023tqo}. With the charge conjugation symmetry, we expect that the magnetic symmetry becomes non-invertible, and $(D-4)$-brane is predicted at any rate. It would be interesting to work out the details.

\section{Nine Dimension \texorpdfstring{$(d=1)$}{d=1}}\label{sec:9d}

In this section, we concretely perform the orbifolding mentioned earlier and see the massless spectrum. First, we will point out the symmetry that the lattice in the 9-dimensional theory possesses and consider orbifolding the theory based on the symmetry. This covers everything that can be constructed from the 9-dimensional SUSY heterotic theory by orbifolding.
To be specific, we focus on outer automorphism and $(-1)^F$ twists of Rank $17$ 9d SUSY heterotic strings. This may correspond to $B_{III}$ theory in Fig.~\ref{fig:theories}. This is because the 10d $E_8$ theory is obtained by the $g(-1)^F$ twist of $E_8\times E_8$ strings, where $g$ exchanges two $E_8$'s. As $B_{III}$ is the compactification $S^1$ of $E_8$, it is likely to be realized by the procedure above. We leave the investigation of other theories for future publications. The result of this section is summarized in Table~\ref{tab:9d_summary} and Table~\ref{tab:9d_gauge_group_summary}.

\begin{table}[h]
    \centering
    \begin{tabular}{|c|c|c|c|c|}
  \hline
  \textbf{SUSY}
  &\cancel{\textbf{SUSY}}
  & \multicolumn{2}{c|}{\textbf{Twisted sector}} \\ \cline{3-4}
  theory& theory
  & $\bar{O}_8$ & $\bar{C}_8$ \\ \hline
  $A_{17}$ & 
  $C_9$  
  & $\mathbf{1}$ & $\mathbf{152}$\\ \hline
  $A_{15}+2A_1$ 
  & $C_8 + A_1$ 
  & $\mathbf{1}$ & $(\mathbf{1},\mathbf{3})\oplus (\mathbf{119},\mathbf{1})$\\ \hline
  $A_{11}+2A_2+2A_1$ 
  & $C_6 + A_2 + A_1$ & $\mathbf{1}$ &  $\qty(\mathbf{65},\mathbf{1},\mathbf{1})\oplus\qty(\mathbf{1},\mathbf{8},\mathbf{1})\oplus\qty(\mathbf{1},\mathbf{1},\mathbf{3})$ 
 \\ \hline
  $A_9+2A_4$ 
  &$C_5+A_4$ & $\mathbf{1}$ &    $ (\mathbf{44},\mathbf{1})\oplus(\mathbf{1},\mathbf{24})$
\\ \hline
  $A_7+2D_5$ 
  &$C_4+D_5$ & $\mathbf{1}$ &
$\qty(\mathbf{27},\mathbf{1})\oplus\qty(\mathbf{1},\mathbf{45})$  \\ \hline
  $A_5+2E_6$ 
  &$C_3+E_6$ & $\mathbf{1}$ & $\qty(\mathbf{14},\mathbf{1})\oplus\qty(\mathbf{1},\mathbf{78})$
\\ \hline
  $A_3+2E_7$ 
  & $C_2+E_7$ 
  & $\mathbf{1}$ & $\qty(\mathbf{5},\mathbf{1})\oplus\qty(\mathbf{1},\mathbf{133})$\\ \hline
  $A_1+2E_8$ 
  & $C_1+E_8$ 
  & $\mathbf{1}$ &    $\qty(\mathbf{1},\mathbf{248})$
 \\ \hline
\end{tabular}
\caption{The list of 9d non-supersymmetric heterotic strings obtained from 9d supersymmetric heterotic strings by orbifolding the outer automorphism and the fermion parity. The untwisted fields are the graviton, B-field, dilaton, gauge fields, gauginos, and the adjoint scalars. The matters coming from the twisted sector are also shown, where $\bar{O}_8$ is the tachyon.}
\label{tab:9d_summary}
\end{table}

\begin{table}[h]
    \centering
    \begin{tabular}{|c|c|c|c|c|}
  \hline
  \textbf{SUSY}
  &\cancel{\textbf{SUSY}}
  \\ \hline
  {\small $\dfrac{\SU(18)\times \U(1)}{\mathbb{Z}_6}\rtimes_I\mathbb{Z}_2$ }& 
  $\dfrac{\Sp(9)\times\U(1)}{\Z_2}$
  \\ \hline
  {\small$\dfrac{\SU(16)\times \SU(2)^2 \times\U(1)}{\Z_8\times\Z_2}\rtimes\Z_2$ }
  & 
  $
    \dfrac{\Sp(8)}{\Z_2}\times\dfrac{\SU(2)\times\U(1)}{\Z_2},$
  \\ \hline
  {\small $
  \dfrac{\SU(3)^2\times\SU(12)\times\SU(2)^2\times\U(1)}{\Z_4\times\Z_3\times\Z_2}\rtimes\Z_2$}
  & 
$\dfrac{\SU(3)\times\Sp(6)\times\SU(2)\times\U(1)}{\Z_3\times\Z_2\times\Z_2}$
 \\ \hline
  $\dfrac{\SU(10)\times\SU(5)^2\times\U(1)}{\Z_{10}\times\Z_5}\rtimes\Z_2$
  & 
  $\dfrac{\SU(5)\times\Sp(5)\times\U(1)}{\Z_2\times\Z_5}$
\\ \hline
  $\dfrac{\SU(8)\times\Spin(10)^2\times\U(1)}{\Z_8\times\Z_4}\rtimes\Z_2$ 
  & 
  $\dfrac{\Sp(4)\times\Spin(10)\times\U(1)}{\Z_2\times\Z_4}$
  \\ \hline
  $\dfrac{\SU(6)\times E_6^2\times \U(1)}{\Z_6\times\Z_3}\rtimes \Z_2$ 
  & 
  $\dfrac{\Sp(3)\times E_6\times\U(1)}{\Z_2\times \Z_3}$
\\ \hline
  $\dfrac{\SU(4)\times E_7^2\times \U(1)}{\Z_4\times\Z_2}\rtimes \Z_2$
  & $\dfrac{\Sp(2)\times E_7\times\U(1)}{\Z_2\times\Z_2}$ 
  \\ \hline
  {\small$\dfrac{\SU(2)\times \U(1)}{\Z_2}\times E_8^2\rtimes\Z_2$}
  & {\small$\dfrac{\SU(2)\times \U(1)}{\Z_2}\times E_8$} 
 \\ \hline
\end{tabular}
\caption{The list of gauge groups of 9d non-supersymmetric and supersymmetric heterotic strings. Here $\U(1)$ corresponds to the graviphoton.}
\label{tab:9d_gauge_group_summary}
\end{table}

\subsection{\texorpdfstring{$A_{17}\to C_9$}{A17->C9}}
The $(17,1)$ even self-dual lattice for $A_{17}$ is given as follows:
\begin{equation}
    \Gamma^{(A_{17})}_{17,1}=\Lambda_R(A_{17})\oplus\Z\sqrt{2}+\Z \qty(6\omega_1^{(A_{17})};0)+\Z\qty(3\omega_1^{(A_{17})};\frac{1}{\sqrt{2}}),
\end{equation}
where $\oplus$ is the direct sum of the abelian groups, and $+$ represents the union of two sets. $\Lambda_R(A_{17})$ is the root lattice of $A_{17}$:

\begin{equation}
    \Lambda_R(A_{17})=\bigoplus_{i=1}^{17}\Z\alpha_i^{(A_{17})}.
\end{equation}
These states are invariant under the $\Z_6$ action $(e^{2\pi i\frac{1}{6}}\mathbf{1},-1)$, where $e^{2\pi i\frac{1}{6}}\mathbf{1}$ is the cubic of the center $\SU(18)$, and the latter $-1$ means the action on the states with a minimal $\U(1)$ charge.
This lattice has the following symmetry:
\begin{equation}
\begin{aligned}
    g:\qty(\sum_{i=1}^{17}x_i \alpha^{(A_{17})}_i;x^{(R)})
    \mapsto\qty(\sum_{i=1}^{17}x_{18-i} \alpha^{(A_{17})}_{i};x^{(R)}).
\end{aligned}
\label{eq:A17_symmetry}\end{equation}
It can be checked as follows:
\begin{equation}
\begin{aligned}
g(\Lambda_R(A_{17})\oplus\Z\sqrt{2})=&\Lambda_R(A_{17})\oplus\Z\sqrt{2},\\
g\qty(\qty(6\omega_1^{(A_{17})};0))=&6\alpha_1^{(A_{17})}+\cdots+6\alpha_{17}^{(A_{17})}-6\omega_1^{(A_{17})},\\
    g\qty(\qty(3\omega_1^{(A_{17})};\frac{1}{\sqrt{2}}))=&\left(3\alpha_1^{(A_{17})}+\cdots+3\alpha_{17}^{(A_{17})}-3\omega_1^{(A_{17})};\frac{1}{\sqrt{2}}\right)\\    =&3\qty(\alpha_1^{(A_{17})}+\cdots+\alpha_{17}^{(A_{17})};0)-\qty(3\omega_1^{(A_{17})};\frac{1}{\sqrt{2}})+\qty(0;\sqrt{2}).
\end{aligned}
\end{equation}

From the argument above, we conclude that the gauge symmetry of supersymmetric theory is
\begin{align}
    \frac{\SU(18)\times \U(1)}{\mathbb{Z}_6}\rtimes_I\mathbb{Z}_2,
\label{eq:A17_gauge_group}\end{align}
where $U(1)$ comes from the graviphoton.
Notice that two possibilities of the semidirect product of the $SU(N)$ gauge group with even $N$ are discussed in \cite{Arias-Tamargo:2019jyh}, dubbed $\widetilde{SU}(N)_I$ and $\widetilde{SU}(N)_{II}$. In our case, the symmetry \eqref{eq:A17_symmetry} maps fundamental to anti-fundamental. The representation matrix is
\begin{align}
    \begin{pmatrix}
    0 & I_9 \\
    I_9 & 0
    \end{pmatrix}_{18 \times 18}.
\end{align}
This corresponds to $\widetilde{SU}(N)_I$, which we write as a subscript in \eqref{eq:A17_gauge_group}.

After the orbifolding, the invariant lattice $I$ and its dual lattice $I^\ast$ are given as follows:
\begin{equation}
\begin{aligned}
        I=&\sqrt{2}\qty(\Lambda_R(C_{9})\oplus\Z1+\Z\qty(\frac{1}{2}\alpha_9^{(C_9)};0)+\qty(\frac{1}{2}\omega_9^{(C_9)};\frac{1}{2})),\\
        I^\ast
        =&\frac{1}{\sqrt{2}}\qty(\Lambda_R(C_{9})\oplus\Z2+\Z\qty(\omega_9^{(C_9)};1))
        =I+\frac{1}{\sqrt{2}}\Lambda_R(C_9).
\end{aligned}
\end{equation}
By gathering the eight neutral elements and  elements $(p_L,p_R)\in I^\ast$ which satisfy $p_L^2=1,p_R^2=0$, we obtain the quasi-minuscule representation (see App.~\ref{sec:lie algebra}) of $C_9$ as massless particles:
\begin{equation}
    \mathbf{152}.
\end{equation}

The charge of the general twisted sector states can be studied from the lattice $I^*$. Similarly, the charge of the untwisted sector states can be read by checking the branching rule of the matter representations under symmetry breaking $A_{17}\to C_9$. Note that we always have $\U(1)$ symmetry corresponding to right moving momentum. The minimal $\U(1)$ charge corresponds to $p_R=1/\sqrt{2}$. The lattice point $p_R=1/\sqrt{2}$ accompanies with $p_L=\omega_9^{(C_9)}/\sqrt{2}$.
Consequently, we find a symmetry of non-supersymmetric theory is\footnote{Other than $(-1)^{f}$ symmetry discussed in Sec.~\ref{sec:fermionization}.}
\begin{align}
    \frac{\Sp(9)\times\U(1)}{\Z_2},
\end{align}
where $\Z_2$ is generated by $\Sp(9)$ center as well as $\pi$ rotation of $\U(1)$.

\subsection{\texorpdfstring{$A_{15}+2A_1\to C_8+A_1$}{A15+2A1->C8+C1}}
The $(17,1)$ even self-dual lattice for $A_{15}+2A_1$ is given as follows:
\begin{equation}
    \Gamma^{(A_{15}+2A_1)}_{17,1}=\Lambda_R(A_{15}+2A_1)\oplus\Z2+\Z \qty(4\omega_1^{(A_{15})},\omega^{(A_1)}_1,\omega_1^{(A_1)};0)+\Z\qty(2\omega^{(A_{15})}_1,0,\omega^{(A_1)}_1;\frac{1}{2}).
\end{equation}
These states are invariant under the action,
\begin{align}
    \Z_8:(e^{2\pi i\frac{1}{8}}\mathbf{1},-\mathbf{1},\mathbf{1},e^{-2\pi i\frac{1}{4}}),
    \quad
    \Z_2:(\mathbf{1},-\mathbf{1},-\mathbf{1},-1).
\end{align}
The lattice has the following symmetry:
\begin{equation}
\begin{aligned}
    g:\qty(\sum_{i=1}^{15}x_i \alpha^{(A_{15})}_i,x_1^{(A_{1})},x^{(A_{1})}_2;x^{(R)})
    \mapsto \qty(\sum_{i=1}^{15}x_{16-i} \alpha^{(A_{15})}_{i},-x^{(A_{1})}_2,-x^{(A_{1})}_1;x^{(R)}).
\end{aligned}
\end{equation}

The invariant lattice $I$ and its dual lattice $I^\ast$ are given as follows:
\begin{equation}
    \begin{aligned}
            I=&\sqrt{2}\qty(\Lambda_R(C_8+A_1)\oplus \Z\sqrt{2}+\Z\qty(\frac{1}{2}\alpha^{(C_8)}_8,0;0)+\Z\qty(\frac{1}{2}\omega_8^{(C_8)},\omega_1^{(A_1)};\frac{1}{\sqrt{2}})),\\
            I^\ast=&\frac{1}{\sqrt{2}}\qty(\Lambda_R(C_8+A_1)\oplus\Z\sqrt{2})
            =I+\frac{1}{\sqrt{2}}\Lambda_R(C_8+A_1)
            .
    \end{aligned}
\end{equation}
By gathering the eight neutral elements and the elements $(p_L,p_R)\in I^\ast$ that satisfy $p_L^2=1,p_R^2=0$, we obtain the following representation of $C_8+A_1$  as massless particles:

\begin{equation}
    (\mathbf{1},\mathbf{3}),~(\mathbf{119},\mathbf{1}),
\end{equation}
where $\mathbf{119}$ is the quasi-minuscule representation of $C_8$ .

The symmetry of non-supersymmetric theory is

\begin{align}
    \frac{\Sp(8)}{\Z_2}\times\frac{\SU(2)\times\U(1)}{\Z_2},
\end{align}

where the second $\Z_2$ is generated by $\SU(2)$ center as well as $\pi$ rotation of $\U(1)$.

\subsection{\texorpdfstring{$A_{11}+2A_2+2A_1\to C_6+A_2+A_1$}{A11+2A2+2A1->C6+A2+A1}}
The $(17,1)$ even self-dual lattice for $A_{11}+2A_2+2A_1$ is given as follows:

\begin{equation}
\begin{aligned}
\Gamma^{(A_{11}+2A_2+2A_1)}_{17,1}=&\Lambda_R(A_{11}+2A_2+2A_1)\oplus\Z\sqrt{12}+\Z\qty(10\omega_1^{(A_{11})},\omega_1^{(A_2)},\omega_1^{(A_2)},\omega_1^{(A_1)},\omega_1^{(A_1)};0)\\
+&\Z\qty(3\omega_1^{(A_{11})},\omega_1^{(A_2)},-\omega_1^{(A_{2})},\omega_1^{(A_1)},0;\frac{1}{\sqrt{12}}).
\end{aligned}
\end{equation}
These states are invariant under the action,
\begin{align}
    &\Z_4:(e^{2\pi i\frac{1}{4}}\mathbf{1},\mathbf{1},\mathbf{1},\mathbf{1},-\mathbf{1},e^{2\pi i\frac{1}{4}}),
    \nonumber\\
    &
    \Z_3:(\mathbf{1},e^{2\pi i\frac{1}{3}}\mathbf{1},e^{-2\pi i\frac{1}{3}}\mathbf{1},\mathbf{1},\mathbf{1},e^{-2\pi i\frac{2}{3}}),
    \nonumber\\
    &\Z_2:(\mathbf{1},\mathbf{1},\mathbf{1},-\mathbf{1},-\mathbf{1},-1).
\end{align}
This lattice has the following symmetry:
\begin{equation}
\begin{aligned}
    g:&\qty(\sum_{i=1}^{11}x_i \alpha^{(A_{11})}_i,x^{(A_{2})}_1,x^{(A_{2})}_2,x^{(A_{1})}_1,x^{(A_{1})}_2;x^{(R)})\\
    \mapsto&\qty(\sum_{i=1}^{11}x_{12-i} \alpha^{(A_{11})}_{i},-x^{(A_{2})}_2,-x^{(A_{2})}_1,-x^{(A_{1})}_2,-x^{(A_{1})}_1;x^{(R)}).
\end{aligned}
\end{equation}
The invariant lattice $I$ and its dual lattice $I^\ast$ are given as follows
\begin{equation}
\begin{aligned}
I=&\sqrt{2}\qty(\Lambda_R(C_6+A_2+A_1)\oplus\Z\sqrt{6}+\Z\frac{1}{2}\alpha_6^{(C_6)}+\Z\qty(\frac{1}{2}\omega_6^{(C_6)},0,\omega_1^{(A_1)};0)+\Z\qty(\frac{1}{2}\omega_6^{(C_6)},2\omega_1^{(A_2)},0;\frac{1}{\sqrt{6}})),\\
    I^\ast=&I+\frac{1}{\sqrt{2}}\Lambda_R(C_6+A_2+A_1).
\end{aligned}
\end{equation}
By gathering the eight neutral elements and  elements $(p_L,p_R)\in I^\ast$ which satisfy $p_L^2=1,p_R^2=0$, we obtain the following representation of $C_6+A_2+A_1$  as massless particles:
\begin{equation}
\qty(\mathbf{65},\mathbf{1},\mathbf{1}),~ \qty(\mathbf{1},\mathbf{8},\mathbf{1}),~ \qty(\mathbf{1},\mathbf{1},\mathbf{3}),
\end{equation}
where $\mathbf{65}$ is the quasi-minuscule representation of $C_6$ , and $\mathbf{8},\mathbf{3}$ are the adjoint representation of  $A_2,A_1$.

The symmetry of the non-supersymmetric theory is

\begin{align}
    \frac{\Sp(6)\times\SU(3)\times\SU(2)\times\U(1)}{\Z_3\times\Z_2\times\Z_2}
\end{align}
where the action of the denominator is
\begin{align}
    &\Z_3:(\mathbf{1},e^{2\pi i\frac{1}{3}}\mathbf{1},\mathbf{1},e^{-2\pi i\frac{2}{3}}),
    \nonumber\\
    &\Z_2:(-\mathbf{1},\mathbf{1},\mathbf{1},-1),
    \nonumber\\
    &\Z_2:(\mathbf{1},\mathbf{1},-\mathbf{1},-1).
\end{align}


\subsection{\texorpdfstring{$A_9+2A_4\to C_5+A_4$}{A9+2A4->C5+A4}}
The $(17,1)$ even self-dual lattice for $A_9+2A_4$ is given as follows:
\begin{equation}
\Gamma^{(A_9+2A_4)}_{17,1}=\Lambda_R(A_9+2A_4)\oplus\Z\sqrt{10}+\Z\qty(4\omega^{(A_9)}_1,\omega_{1}^{(A_4)},\omega^{(A_4)}_1;0)+\Z\qty(3\omega^{(A_9)}_1,\omega_{1}^{(A_4)},3\omega^{(A_4)}_1;\frac{1}{\sqrt{10}}).
\end{equation}
These states are invariant under the action,
\begin{align}
    \Z_{10}:(e^{2\pi i\frac{1}{10}}\mathbf{1},\mathbf{1},e^{-2\pi i\frac{2}{5}}\mathbf{1},e^{2\pi i\frac{9}{10}}),
    \quad
    \Z_5:(\mathbf{1},e^{2\pi i\frac{1}{5}}\mathbf{1},e^{-2\pi i\frac{1}{5}}\mathbf{1},e^{2\pi i\frac{2}{5}}).
\end{align}
This lattice has the following symmetry:
\begin{equation}
\begin{aligned}
    g:\qty(\sum_{i=1}^{9}x_i \alpha^{(A_{9})}_i,x^{(A_{4})}_1,x^{(A_{4})}_2;x^{(R)})
    \mapsto \qty(\sum_{i=1}^{9}x_{10-i} \alpha^{(A_{9})}_{i},-x^{(A_{4})}_2,-x^{(A_{4})}_1;x^{(R)})
\end{aligned}
\end{equation}
The invariant lattice $I$ and its dual lattice  $I^\ast$ are given as follows:
\begin{equation}
\begin{aligned}
    I=&\sqrt{2}\qty( \Lambda_R(C_5+A_4)\oplus\sqrt{5}+\Z \qty(\frac{1}{2}\alpha_5^{(C_5)},0;0)+\Z\frac{1}{2}\qty(\omega_5^{(C_5)},-2\omega_1^{(A_4)};\frac{1}{\sqrt{5}})),\\
    I^\ast=&I+\frac{1}{\sqrt{2}}\Lambda_R(C_5+A_4).
\end{aligned}
\end{equation}
By gathering the eight neutral elements and  elements $(p_L,p_R)\in I^\ast$ which satisfy $p_L^2=1,p_R^2=0$, we obtain the following representation of $C_5+A_4$  as massless particles:
\begin{equation}
    (\mathbf{44},\mathbf{1}),~ (\mathbf{1},\mathbf{24}),
\end{equation}
where $\mathbf{44}$ is the quasi-minuscule representation of $C_5$ , and $\mathbf{24}$ is the adjoint representation of $A_4$.
The symmetry of non-supersymmetric theory is

\begin{align}
    \frac{\SU(5)\times\Sp(5)\times\U(1)}{\Z_5\times\Z_2},
\end{align}
where $\Z_5$ is generated by $\SU(5)$ center and $4\pi/5$ rotation of $\U(1)$ center,

and $\Z_2$ is generated by $\Sp(5)$ center as well as $\pi$ rotation of $\U(1)$.

\subsection{\texorpdfstring{$A_7+2D_5\to C_4+D_5$}{A7+2D5->C4+D5}}
The $(17,1)$ even self-dual lattice for $A_7+D_5$ is given as follows: 
\begin{equation}
\begin{aligned}
    \Gamma^{(A_7+2D_5)}_{17,1}=&\Lambda_R(A_7+2D_5)\oplus \Z\sqrt{8}+\Z \qty(2\omega_1^{(A_7)},\omega_5^{(D_5)},\omega_5^{(D_5)};0)\\
+&\Z\qty(3\omega^{(A_7)}_1,\omega^{(D_5)}_5,2\omega^{(D_5)}_5;\frac{1}{\sqrt{8}}).
\end{aligned}
\end{equation}
These states are invariant under the action,
\begin{align}
    \Z_{8}:(e^{2\pi i\frac{1}{8}}\mathbf{1},\mathbf{1},e^{-2\pi i\frac{1}{4}}\mathbf{1},e^{2\pi i\frac{1}{8}}),
    \quad
    \Z_4:(\mathbf{1},e^{2\pi i\frac{1}{4}}\mathbf{1},e^{-2\pi i\frac{1}{4}}\mathbf{1},e^{2\pi i\frac{1}{4}}).
\end{align}
This lattice has the following symmetry:
\begin{equation}
\begin{aligned}
    g:\qty(\sum_{i=1}^{7}x_i \alpha^{(A_{7})}_i,x^{(D_{5})}_1,x^{(D_{5})}_2;x^{(R)})
    \mapsto \qty(\sum_{i=1}^{7}x_{8-i} \alpha^{(A_{7})}_{i},-x^{(D_{5})}_2,-x^{(D_{5})}_1;x^{(R)}).
\end{aligned}
\end{equation}
The invariant lattice $I$ and its dual lattice  $I^\ast$ are
\begin{equation}
\begin{aligned}
    I=&\sqrt{2}\qty(\Lambda_R(C_4+D_5)\oplus\Z2+\Z\qty(\frac{1}{2}\alpha_4^{(C_4)},0;0)+\Z\qty(\frac{1}{2}\omega_4^{(C_4)},2\omega_5^{(D_5)};0)+\Z\qty(0,\omega_5^{(D_5)};-\frac{1}{2})),\\
    I^\ast=&I+\frac{1}{\sqrt{2}}\Lambda_R(C_4+D_5).
\end{aligned}
\end{equation}
By gathering the eight neutral elements and  elements $(p_L,p_R)\in I^\ast$ which satisfy $p_L^2=1,p_R^2=0$, we obtain the following representation of $C_4+D_5$  as massless particles:
\begin{equation}
\qty(\mathbf{27},\mathbf{1}),~\qty(\mathbf{1},\mathbf{45}), 
\end{equation}
where $\mathbf{27}$ is the quasi-minuscule representation of $C_4$, and $\mathbf{45}$ is the adjoint representation of  $D_5$. 
The symmetry of non-supersymmetric theory is

\begin{align}    \frac{\Sp(4)\times\Spin(10)\times\U(1)}{\Z_2\times\Z_4},
\end{align}
where $\Z_2$ is $\Sp(4)$ center as well as $\pi$ rotation of $\U(1)$, and $\Z_4$ is generated by ($\Spin(10)$ center) as well as $\pi/2$ rotation of $\U(1)$.

\subsection{\texorpdfstring{$A_5+2E_6\to C_3+E_6$}{A5+2E6->C3+E6}}

The $(17,1)$ even self-dual lattice for $A_5+2E_6$ is given as follows:
\begin{equation}
\Gamma^{(A_5+2E_6)}_{17,1}=\Lambda_R(A_5+2E_6)\oplus\Z\sqrt{6}+\Z\qty(2\omega_1^{(A_5)},\omega_6^{(E_6)},\omega_6^{(E_6)};0)+\Z\qty(\omega_1^{(A_5)},0,\omega_6^{(E_6)};\frac{1}{\sqrt{6}}).
\end{equation}
These states are invariant under the action,
\begin{align}
    \Z_{6}:(e^{2\pi i\frac{1}{6}}\mathbf{1},\mathbf{1},e^{-2\pi i\frac{1}{3}}\mathbf{1},e^{2\pi i\frac{1}{6}}),
    \quad
    \Z_3:(\mathbf{1},e^{2\pi i\frac{1}{3}}\mathbf{1},e^{-2\pi i\frac{1}{3}}\mathbf{1},e^{2\pi i\frac{1}{3}}).
\end{align}
This lattice has the following symmetry:
\begin{equation}
\begin{aligned}
    g:\qty(\sum_{i=1}^{5}x_i \alpha^{(A_{5})}_i,x^{(E_6)}_1,x^{(E_6)}_2;x^{(R)})\mapsto \qty(\sum_{i=1}^{5}x_{6-i} \alpha^{(A_{5})}_{i},-x^{(E_6)}_2,-x^{(E_6)}_1;x^{(R)}).
\end{aligned}
\end{equation}
The invariant lattice $I$ and its dual lattice  $I^\ast$ are
\begin{equation}
    \begin{aligned}
    I=&\sqrt{2}\qty(\Lambda_R(C_3+E_6)\oplus\Z\sqrt{3}+\Z\qty(\frac{1}{2}\alpha_3^{(C_3)},0;0)+\Z\frac{1}{2}\qty(\omega_3^{(C_3)},2\omega_6^{(E_6)};\frac{1}{\sqrt{3}})),\\
    I^\ast=&I+\frac{1}{\sqrt{2}}\Lambda_R(C_3+E_6).
    \end{aligned}
\end{equation}
By gathering the eight neutral elements and  elements $(p_L,p_R)\in I^\ast$ which satisfy $p_L^2=1,p_R^2=0$, we obtain the following representation of $C_3+E_6$:
\begin{equation}
    \qty(\mathbf{14},\mathbf{1}),~\qty(\mathbf{1},\mathbf{78}),
\end{equation}
where $\mathbf{14}$ is the quasi-minuscule representation of $C_3$ and $\mathbf{78}$ is the adjoint representation of $E_6$.
The symmetry of non-supersymmetric theory is

\begin{align}
    \frac{\Sp(3)\times E_6\times\U(1)}{\Z_3\times \Z_2},
\end{align}

where the $\Z_3$ is generated by $E_6$ center as well as $2\pi/3$ rotation of $\U(1)$, and $\Z_2$ is generated by $\Sp(3)$ center as well as $\pi$ rotation of $\U(1)$.

\subsection{\texorpdfstring{$A_3+2E_7\to C_2+E_7$}{A3+2E7->C2+E7}}

The $(17,1)$ even self-dual lattice for $A_3+2E_7$ is given as follows:
\begin{equation}
\Gamma^{(A_3+2E_7)}_{17,1}=\Lambda_R(A_3+2E_7)\oplus\Z2+\Z\qty(2\omega_1^{(A_3)},\omega_7^{(E_7)},\omega_7^{(E_7)};0)+\Z\qty(\omega_1^{(A_3)},0,\omega_7^{(E_7)};\frac{1}{2}).
\end{equation}
These states are invariant under the action,
\begin{align}
    \Z_4:(e^{2\pi i\frac{1}{4}}\mathbf{1},\mathbf{1},-\mathbf{1},e^{2\pi i\frac{1}{4}}),
    \quad
    \Z_2:(\mathbf{1},-\mathbf{1},-\mathbf{1},-1).
\end{align}
This lattice has the following symmetry:
\begin{equation}
\begin{aligned}
    g:\qty(\sum_{i=1}^{3}x_i \alpha^{(A_{3})}_i,x^{(E_7)}_1,x_2^{(E_7)};x^{(R)})
    \mapsto \qty(\sum_{i=1}^{3}x_{4-i} \alpha^{(A_{3})}_{i},-x^{(E_7)}_2,-x^{(E_7)}_1;x^{(R)}).
\end{aligned}
\end{equation}
The invariant lattice $I$ and its dual lattice  $I^\ast$ are
\begin{equation}
\begin{aligned}
I=&\sqrt{2}\qty(\Lambda_R(C_2+E_7)\oplus\Z\sqrt{2}+\Z\left(\frac{1}{2}\alpha_2^{(C_2)},0;0\right)+\Z\qty(\frac{1}{2}\omega_2^{(C_2)},\omega_7^{(E_7)};0)+\Z\qty(\frac{1}{2}\omega_2^{(C_2)},0;\frac{1}{\sqrt{2}})),\\
    I^\ast=&I+\frac{1}{\sqrt{2}}\Lambda_R(C_2+E_7).
\end{aligned}
\end{equation}
By gathering the eight neutral elements and  elements $(p_L,p_R)\in I^\ast$ which satisfy $p_L^2=1,p_R^2=0$ ,we obtain the following representation of $C_2+E_7$  as massless particles:
\begin{equation}
\qty(\mathbf{5},\mathbf{1}),~\qty(\mathbf{1},\mathbf{133}),
\end{equation}
where $\mathbf{5}$ is the quasi-minuscule representation of $C_2$ , and $\mathbf{133}$ is the adjoint representation of  $E_7$.

The symmetry of non-supersymmetric theory is
\begin{align}
    \frac{\Sp(2)\times E_7\times\U(1)}{\Z_2\times\Z_2},
\end{align}
where the first $\Z_2$ is ($\Sp(2)$ center)$\times$($E_7$ center) while the latter $\Z_2$ is $E_7$ center as well as $\pi$ rotation of $\U(1)$.

\subsection{\texorpdfstring{$A_1+2E_8\to C_1+E_8$}{A1+2E8->C1+E8}}

The $(17,1)$ even self-dual lattice for $A_1+2E_8$ is given as follows:
\begin{equation}
\Gamma^{(A_1+2E_8)}_{17,1}=\Lambda_R(A_1+2E_8)\oplus\Z\sqrt{2}+\Z\qty(\omega_1^{(A_1)},0,0;\frac{1}{\sqrt{2}}).
\end{equation}
These states are invariant under the $\Z_2$ action $(-\mathbf{1},\mathbf{1},\mathbf{1},-1)$.
This lattice has the following symmetry: 
\begin{equation}
\begin{aligned}
g:\qty(x^{(A_1)},x^{(E_8)}_1,x^{(E_8)}_2;x^{(R)})
    \mapsto &\qty(x^{(A_1)},-x^{(E_8)}_2,-x^{(E_8)}_1;x^{(R)})
\end{aligned}
\end{equation}
The invariant lattice $I$ and its dual lattice  $I^\ast$ are
\begin{equation}
\begin{aligned}
        I=&\sqrt{2}\qty(\Lambda_R(C_1+E_8)\oplus\Z1+\Z\qty(\frac{1}{2}\alpha^{(C_1)}_1,0;0)+\Z\qty(\frac{1}{2}\omega_1^{(C_1)},0;\frac{1}{2})),\\
        I^\ast=&I+\frac{1}{\sqrt{2}}\Lambda_R(C_1+E_8).
\end{aligned}
\end{equation}
Of course $C_1$ is isomorphic to $A_1$ as a lie algebra, but their roots have different length($\qty(\alpha_1^{(A_1)})^2=2$, and $\qty(\alpha_1^{(C_1)})^2=4$). 

By gathering the eight neutral elements and the elements $(p_L,p_R)\in I^\ast$ that satisfy $p_L^2=1,p_R^2=0$, we obtain the following representation of $C_1+E_8$  as massless particles:
\begin{equation}
    \qty(\mathbf{1},\mathbf{248}),
\end{equation}
where $\mathbf{248}$ is the adjoint representation of  $E_8$.

\subsection{Impossible Folding}

Interestingly, there is a nice bottom-up argument why only the folding $A_{2n-1}\to C_n$ appears in this case. For instance, Suppose that there exists the gauge group of the form $\Spin(2n)\rtimes\Z_2$. Then, we can perform the twist compactification on $S^1$ such that all fields are twisted by $\Z_2$. Consequently, we will obtain an eight-dimensional supersymmetric theory with $B_{n-1}$ symmetry. However, this contradicts with global anomaly of 8d supergravity~\cite{Garcia-Etxebarria:2017crf}. Similarly, the disconnected gauge group leading to $B_n$ and $F_4$ are prohibited by the global anomaly. Furthermore, the group of the form $\Spin(8)\rtimes S_3$ leads to $G_2$ gauge symmetry in 8d. This is not possible due to an argument based on the brane probe~\cite{Hamada:2021bbz,Bedroya:2021fbu}. To summarize, as a non-simply-laced group, we only obtain $C_n$ as a 9d non-supersymmetric heterotic string theory.

\subsection{Relation with 9d and 8d CHL strings}

The eight gauge symmetries we have identified have relationships to the CHL strings that should be explained. In 9d and 8d CHL strings, there are gauge symmetries that look similar to our eight symmetries~\cite{Font:2021uyw}, as shown in Table~\ref{tab:nonSUSY_vs_CHL}.

\begin{table}[h]
\centering
\begin{tabular}{|c|c|c|}
\hline
9d CHL & 9d \cancel{\textbf{SUSY}} & 8d CHL \\
\hline
$A_9$ & $C_9$ & $C_9 + A_1$ \\ \hline
$A_8 + A_1$ & $C_8 + A_1$ & $C_8 + A_1 + A_1$ \\ \hline
$A_6 + A_2 + A_1$ & $C_6 + A_2 + A_1$ & $C_6 + A_2 + A_1 + A_1$ \\ \hline
$A_5 + A_4$ & $C_5 + A_4$ & $C_5 + A_4 + A_1$ \\ \hline
$A_4 + D_5$ & $C_4 + D_5$ & $C_4 + D_5 + A_1$ \\ \hline
$A_3 + E_6$ & $C_3 + E_6$ & $C_3 + E_6 + A_1$ \\ \hline
$A_2 + E_7$ & $C_2 + E_7$ & $C_2 + E_7 + A_1$ \\ \hline
$A_1 + E_8$ & $C_1 + E_8$ & $C_1 + E_8 + A_1$ \\ \hline
\end{tabular}
\caption{The symmetries we have identified and a part of the symmetries of the CHL strings. These symmetries are similar.}
    \label{tab:nonSUSY_vs_CHL}
\end{table}

The relationships are given as follows:
\begin{equation}
    \begin{aligned}
        &\text{9d } \cancel{\textbf{SUSY}}~\overset{C_n\to A_n}{\longrightarrow} &&\text{9d CHL},\\
        &\text{9d } \cancel{\textbf{SUSY}}~~~\overset{+A_1}{\longrightarrow} &&\text{8d CHL}.\\
    \end{aligned}
\end{equation}
We plan to work on explaining these relationships in the future.

\section{Eight Dimension \texorpdfstring{$(d=2)$}{(d=2)}}\label{sec:8d}

In this section, we consider several examples of 8d non-supersymmetric heterotic strings constructed from supersymmetric strings.
All eight $(17,1)$ lattices  which were orbifolded in previous section appear in eight dimensions as follows:
\begin{equation}
    \Gamma_{18,2}=\Gamma_{17,1}\oplus \Gamma_{1,1},
\end{equation}
where $\Gamma_{1,1}$ is the $(1,1)$ even self-dual lattice:
\begin{equation}
\Gamma_{1,1}=\Lambda_R(A_1)\oplus\Z\sqrt{2}+\Z\qty(\omega_1^{(A_1)};\frac{1}{\sqrt{2}}).
\end{equation}

Invariant lattice can be obtained as well:
\begin{equation}
\begin{aligned}
        I_{10,2}=&I_{9,1}\oplus \sqrt{2}\qty(\Lambda_R(C_1)\oplus\Z1+\Z\frac{1}{2}\alpha_1^{(C_1)}),\\
        I_{10,2}^\ast=&I_{9,1}^\ast\oplus \frac{1}{\sqrt{2}}\qty(\Lambda_R(C_1)\oplus\Z2+\Z\omega_1^{(C_1)}).
\end{aligned}
\end{equation}

Of course $C_1$ and $A_1$ are isomorphic as a lie algebra, but their roots and weights have different length:
\begin{equation}
    \begin{aligned}
        \alpha_1^{(C_1)}=\sqrt{2}\alpha_1^{(A_1)},\\
        \omega_1^{(C_1)}=\sqrt{2}\omega_1^{(A_1)}.
    \end{aligned}
\end{equation}
In Table~\ref{tab:8d_list}, we list 8d theories obtained in this way.

\begin{table}[h]
    \centering
    \begin{tabular}{|c|c|c|c|c|}
  \hline
  \textbf{SUSY}
  &\cancel{\textbf{SUSY}}
  & \multicolumn{2}{c|}{\textbf{Twisted sector}} \\ \cline{3-4}
  theory& theory
  & $\bar{O}_8$ & $\bar{C}_8$ \\ \hline
  $A_{17}+A_1$ & 
  $C_9+C_1$  
  & $\mathbf{1}$ & $(\mathbf{152},\mathbf{1})$\\ \hline
  $A_{15}+3A_1$ 
  & $C_8 +C_1+ A_1$ 
  & $\mathbf{1}$ & $(\mathbf{1},\mathbf{1},\mathbf{3})\oplus (\mathbf{119},\mathbf{1},\mathbf{1})$\\ \hline
  $A_{11}+2A_2+3A_1$ 
  & $C_6 +C_1 + A_2 + A_1$ & $\mathbf{1}$ &  $\qty(\mathbf{65},\mathbf{1},\mathbf{1},\mathbf{1})\oplus\qty(\mathbf{1},\mathbf{1},\mathbf{8},\mathbf{1})\oplus\qty(\mathbf{1},\mathbf{1},\mathbf{1},\mathbf{3})$ 
 \\ \hline
  $A_9+2A_4+A_1$ 
  &$C_5+C_1+A_4$ & $\mathbf{1}$ &    $ (\mathbf{44},\mathbf{1},\mathbf{1})\oplus(\mathbf{1},\mathbf{1},\mathbf{24})$
\\ \hline
  $A_7+2D_5+A_1$ 
  &$C_4+C_1+D_5$ & $\mathbf{1}$ &
$\qty(\mathbf{27},\mathbf{1},\mathbf{1})\oplus\qty(\mathbf{1},\mathbf{45})$  \\ \hline
  $A_5+2E_6+A_1$ 
  &$C_3+C_1+E_6$ & $\mathbf{1}$ & $\qty(\mathbf{14},\mathbf{1},\mathbf{1})\oplus\qty(\mathbf{1},\mathbf{1},\mathbf{78})$
\\ \hline
  $A_3+2E_7+A_1$ 
  & $C_2+C_1+E_7$ 
  & $\mathbf{1}$ & $\qty(\mathbf{5},\mathbf{1},\mathbf{1})\oplus\qty(\mathbf{1},\mathbf{1},\mathbf{133})$\\ \hline
  $2A_1+2E_8$ 
  & $2C_1+E_8$ 
  & $\mathbf{1}$ &    $\qty(\mathbf{1},\mathbf{1},\mathbf{248})$
 \\ \hline
\end{tabular}
\caption{The list of 8d non-supersymmetric and supersymmetric heterotic strings coming from $S^1$ reduction of 9d strings.}
\label{tab:8d_list}\end{table}

On top of these theories, there are 8d heterotic strings, which can not be obtained by adding $C_1$ gauge algebra to 9d gauge algebra. We do not provide a complete list, but we give two examples in this section (see Table~\ref{tab:8d_list_2}). We leave the comprehensive analysis for future publications.

\begin{table}[h]
    \centering
    \begin{tabular}{|c|c|c|c|c|}
  \hline
  \textbf{SUSY}
  &\cancel{\textbf{SUSY}}
  & \multicolumn{2}{c|}{\textbf{Twisted sector}} \\ \cline{3-4}
  theory& theory
  & $\bar{O}_8$ & $\bar{C}_8$ \\ \hline
  $A_3+3A_5$ & 
  $C_2+C_3+A_5$  
 &$\mathbf{1}$ & $\qty(\mathbf{5},\mathbf{1},\mathbf{1}),~ \qty(\mathbf{1},\mathbf{14},\mathbf{1}), ~ \qty(\mathbf{1},\mathbf{1},\mathbf{24})$\\ \hline
  $A_{15}+A_3$ 
  & $C_8 +C_2$ 
  & $\mathbf{1}$ & $\qty(\mathbf{119},\mathbf{1}),~\qty(\mathbf{1},\mathbf{5})$\\ \hline
\end{tabular}
\caption{Other examples of 8d non-supersymmetric and supersymmetric heterotic strings.}
\label{tab:8d_list_2}\end{table}

\subsection{\texorpdfstring{$A_3+3A_5\to C_2+C_3+A_5$}{A3+3A5->C2+C3+A5}}
We start from the lattice
\begin{equation}
    \begin{aligned}
\Gamma^{(A_3+3A_5)}_{18,2}=&\Lambda_R(A_3+3A_5)\oplus\Z2\oplus\Z\sqrt{6}+\Z\qty(2\omega_1^{(A_3)},2\omega_1^{(A_5)},5\omega_1^{(A_5)},5\omega_1^{(A_5)};0,0)\\
    +&\Z\qty(\omega_1^{(A_3)},0,3\omega_1^{(A_5)},0;\frac{1}{2},0)+\Z\qty(0,\omega_1^{(A_5)},2\omega_1^{(A_5)},0;0,\frac{1}{\sqrt{6}}).
    \end{aligned}
\end{equation}
This lattice has the following symmetry:
\begin{equation}
    \begin{aligned}
g:&\qty(\sum_{i=1}^3 x_i \alpha_i^{(A_3)},\sum_{i=1}^5 x_i \alpha_i^{(A_5)},x^{(A_5)}_1,x^{(A_5)}_2;x_1^{(R)},x_2^{(R)})\\
    \mapsto&\qty(\sum_{i=1}^3 x_{4-i} \alpha_i^{(A_3)},\sum_{i=1}^5 x_{6-i} \alpha_i^{(A_5)},-x^{(A_5)}_2,-x^{(A_5)}_1;x_1^{(R)},x_2^{(R)}).
    \end{aligned}
\end{equation}
The invariant lattice $I$ and its dual lattice  $I^\ast$ are
\begin{equation}
    \begin{aligned}
I=&\sqrt{2}\Biggl(\Lambda_R(C_2+C_3+A_5)\oplus\Z\sqrt{2}\oplus\Z\sqrt{3}+\Z\qty(\frac{1}{2}\alpha_2^{(C_2)},0;0,0)+\Z\qty(0,\frac{1}{2}\alpha_3^{(C_3)};0,0)\\
+&\Z\qty(\frac{1}{2}\omega_2^{(C_2)},0,3\omega_1^{(A_5)};0,0) 
    +\Z\qty(\frac{1}{2}\omega_2^{(C_2)},0,0,0;\frac{1}{\sqrt{2}},0)+\Z\qty(0,\frac{1}{2}\omega_3^{(C_3)},0;0,\frac{\sqrt{3}}{2})\Biggr),\\
    I^\ast=&I+\frac{1}{\sqrt{2}}\Lambda_R(C_2+C_3+A_5).
    \end{aligned}
\end{equation}

By gathering the eight neutral elements and elements $(p_L,p_R)\in I^\ast$ that satisfy $p_L^2=1, p_R^2=0$, we obtain the following representation of $C_2+C_3+A_5$  as massless particles:
\begin{equation}
    \qty(\mathbf{5},\mathbf{1},\mathbf{1}),~ \qty(\mathbf{1},\mathbf{14},\mathbf{1}), ~ \qty(\mathbf{1},\mathbf{1},\mathbf{24}).
\end{equation}

\subsection{\texorpdfstring{$A_{15}+A_3\to C_8+C_2$}{A15+A3->C8+C2}}
We start from the lattice
\begin{equation}
    \begin{aligned}
\Gamma_{18,2}^{(A_{15}+A_3)}=&\Lambda_R(A_{15}+A_3)\oplus\Z\sqrt{2}\oplus\Z\sqrt{2}+\Z\qty(4\omega_1^{(A_{15})},2\omega_1^{(A_3)};0,0)\\
+&\Z\qty(2\omega_1^{(A_{15})},\omega_1^{(A_3)};\frac{1}{\sqrt{2}},0)+\Z\qty(2\omega_1^{(A_{15})},-\omega_1^{(A_3)};0,\frac{1}{\sqrt{2}}).
    \end{aligned}
\end{equation}
This lattice has the following symmetry:
\begin{equation}
    \begin{aligned}
        g:\qty(\sum_{i=1}^{15} x_i \alpha_i^{(A_{15})},\sum_{i=1}^3 x_i \alpha_i^{(A_3)};x_1^{(R)},x_2^{(R)})\to \qty(\sum_{i=1}^{15} x_{16-i} \alpha_i^{(A_{15})},\sum_{i=1}^3 x_{4-i} \alpha_i^{(A_3)};x_1^{(R)},x_2^{(R)}).
    \end{aligned}
\end{equation}
The invariant lattice $I$ and its dual lattice  $I^\ast$ are
\begin{equation}
    \begin{aligned}
I=&\sqrt{2}\Biggl(\Lambda_R(C_8+C_2)\oplus\Z1\oplus\Z1+\Z\qty(\frac{1}{2}\alpha_8^{(C_8)},0;0,0)+\Z\qty(0,\frac{1}{2}\alpha_2^{(C_2)};0,0)\\
+&\Z\qty(\frac{1}{2}\omega_8^{(C_8)},0;0,0)+\Z\qty(0,\frac{1}{2}\omega_2^{(C_2)};\frac{1}{2},\frac{1}{2})\Biggr),\\
I^\ast=&I+\frac{1}{\sqrt{2}}\Lambda_R(C_8+C_2).
    \end{aligned}
\end{equation}

By gathering the eight neutral elements and the elements $(p_L,p_R)\in I^\ast$ that satisfy $p_L^2=1,p_R^2=0$, we obtain the following representation of $C_8+C_2$  as massless particles:
\begin{equation}
    \qty(\mathbf{119},\mathbf{1}),~\qty(\mathbf{1},\mathbf{5}).
\end{equation}

\section{Discussions and Future Directions}\label{sec:future}

We have seen that the charge lattices of 9d supersymmetric heterotic strings admit $\Z_2$ outer automorphism symmetry for several cases. 
By orbifolding the theory with $\Z_2$ and the fermion parity, we have obtained the eight maximal gauge enhancements in the non-supersymmetric theory listed in Table~\ref{tab:9d_summary}. This theory is expected to belong to the branch of $E_8$ string on $S^1$.
Curiously, these eight gauge symmetries resemble maximal gauge enhancements in the 9d CHL string. Concretely speaking, the gauge symmetries in Table~\ref{tab:9d_summary} are obtained by replacing $A_n$ with $C_n$ in Table 3 of \cite{Font:2021uyw} other than the $D_9$ symmetry which does not appear in our list.

At the same time, various 6-branes that generate $\Z_2$ as a holonomy are predicted. A 6-brane corresponding to $A_1 + E_8 + E_8\to C_1 + E_8$ is viewed as a simple reduction of the heterotic 7-brane~\cite{Kaidi:2023tqo}. On the other hand, for other cases, the exchange $\Z_2$ is partially or fully mutated to charge conjugation symmetry of $\SU(N)$ gauge symmetry. It is interesting to study the theory describing the near horizon geometry of these 6-branes.

It is also interesting to perform the same analysis for other components of the 9d moduli space, namely $A_I, B_{IIa}, B_{IIb}$ theories. For instance, the all gauge group and matter contents are likely to be obtained by twisting $(-1)^F$ and inner automorphisms of the Rank $17$ theory.

Another area to study is the lower dimensions. Furthermore, starting at 8d, the partition function of non-supersymmetric heterotic string theory may not be the form of \eqref{eq:non-SUSY_partition_function} since there is no reason to maintain $D_4$ symmetry for right-moving fermions. It is interesting to study the corresponding supersymmetric theory.

It is desirable to understand this work in the context of duality. One challenging idea is to explore similar symmetries on the F-theory side and perform a similar orbifolding, thereby extending the Het/F-theory duality~\cite{Vafa:1996xn,Morrison:1996na,Morrison:1996pp} to the non-supersymmetric case.

Furthermore, in recent years, the general cohomology theory known as Topological Modular Forms (TMF) has been applied to the study of heterotic string theory~\cite{Tachikawa:2021mby,Tachikawa:2021mvw}, showing intriguing connections with non-supersymmetric heterotic strings ~\cite{Tachikawa:2024ucm,Saxena:2024eil}.

\acknowledgments
We thank Yugo Takanashi for teaching us foldings of Dynkin diagrams.
The work of Y.H. was supported by MEXT Leading Initiative for Excellent Young Researchers Grant No.JPMXS0320210099, JSPS KAKENHI Grants No.24H00976 and 24K07035.

\newpage
\begin{appendix}

\section{Theta functions and \texorpdfstring{$D_4$}{D4} characters}\label{sec:theta}
In this appendix, we summarize the formula used in the paper.
Let $\tau$ be a complex number with positive imaginary part, and  $q=\exp 2\pi i \tau$. Theta functions are defined as follows:

\begin{equation}
\begin{aligned}
& \theta_1(\tau)\coloneqq i \sum_{n \in \mathbb{Z}}(-1)^n q^{\frac{1}{2}\left(n-\frac{1}{2}\right)^2}=0, \\
& \theta_2(\tau)\coloneqq\sum_{n \in \mathbb{Z}} q^{\frac{1}{2}\left(n-\frac{1}{2}\right)^2},  \\
& \theta_3(\tau)\coloneqq\sum_{n \in \mathbb{Z}} q^{\frac{1}{2} n^2}, \\
& \theta_4(\tau)\coloneqq\sum_{n \in \mathbb{Z}}(-1)^n q^{\frac{1}{2} n^2}.
\end{aligned}
\end{equation}
They can also be expressed as infinite products:
\begin{equation}
\begin{aligned}
& \theta_2(\tau)=2 q^{\frac{1}{8}} \prod_{m=1}^{\infty}\left(1-q^m\right)\left(1+ q^m\right)\left(1+ q^m\right), \\
& \theta_3(\tau)=\prod_{m=1}^{\infty}\left(1-q^m\right)\left(1+ q^{m-\frac{1}{2}}\right)\left(1+ q^{m-\frac{1}{2}}\right), \\
& \theta_4(\tau)=\prod_{n=0}^{\infty}\left(1-q^m\right)\left(1- q^{m-\frac{1}{2}}\right)\left(1-q^{m-\frac{1}{2}}\right) .
\end{aligned}
\end{equation}
Modular transformation properties of theta functions are given as follows:
\begin{equation}\label{eq:theta}
    \begin{aligned}
        \theta_2(\tau+1)=&e^{\frac{\pi i}{4}}\theta_2(\tau),\\
        \theta_3(\tau+1)=&\theta_4(\tau),\\
        \theta_4(\tau+1)=&\theta_3(\tau),\\
        \theta_2\qty(-\frac{1}{\tau})=&e^{-\frac{i\pi}{4}}\tau^{\frac{1}{2}}\theta_4(\tau),\\
        \theta_3\qty(-\frac{1}{\tau})=&e^{-\frac{i\pi}{4}}\tau^{\frac{1}{2}}\theta_3(\tau),\\
        \theta_4\qty(-\frac{1}{\tau})=&e^{-\frac{i\pi}{4}}\tau^{\frac{1}{2}}\theta_2(\tau).
    \end{aligned}
\end{equation}
The trivial conjugacy class (the root lattice):
\begin{equation}
    \Gamma_g^{(4)}=\left\{\left(n_1, \cdots, n_4\right) \mid n_i \in \mathbb{Z}, \sum_{i=1}^4 n_i \in 2 \mathbb{Z}\right\} .
\end{equation}
The vector conjugacy class:
\begin{equation}
    \Gamma_v^{(4)}=\left\{\left(n_1, \cdots, n_4\right) \mid n_i \in \mathbb{Z}, \sum_{i=1}^4 n_i \in 2 \mathbb{Z}+1\right\} .
\end{equation}
The spinor conjugacy class:

\begin{equation}
    \Gamma_s^{(4)}=\left\{\left.\left(n_1+\frac{1}{2}, \cdots, n_4+\frac{1}{2}\right) \right\rvert\, n_i \in \mathbb{Z}, \sum_{i=1}^4 n_i \in 2 \mathbb{Z}\right\} .
\end{equation}
The conjugate spinor conjugacy class:

\begin{equation}
    \Gamma_c^{(4)}=\left\{\left.\left(n_1+\frac{1}{2}, \cdots, n_4+\frac{1}{2}\right) \right\rvert\, n_i \in \mathbb{Z}, \sum_{i=1}^4 n_i \in 2 \mathbb{Z}+1\right\}.
\end{equation}
These lattices give the following functions:
\begin{equation}
\begin{aligned}
& O_{8}=\frac{1}{\eta^4} \sum_{p\in \Gamma_g^{(4)}} q^{\frac{1}{2}|\pi|^2}=\frac{1}{2 \eta^4}\left(\theta_3^4(\tau)+\theta_4^4(\tau)\right), \\
& V_{8}=\frac{1}{\eta^4} \sum_{p\in \Gamma_v^{(4)}} q^{\frac{1}{2}|\pi|^2}=\frac{1}{2 \eta^4}\left(\theta_3^4(\tau)-\theta_4^4(\tau)\right), \\
& S_{8}=\frac{1}{\eta^4} \sum_{p\in \Gamma_s^{(4)}} q^{\frac{1}{2}|\pi|^2}=\frac{1}{2 \eta^4}\left(\theta_2^4(\tau)+\theta_1^4(\tau)\right), \\
& C_{8}=\frac{1}{\eta^4} \sum_{p\in \Gamma_c^{(4)}} q^{\frac{1}{2}|\pi|^2}=\frac{1}{2 \eta^4}\left(\theta_2^4(\tau)-\theta_1^4(\tau)\right).
\label{eq:D4_characters}\end{aligned}
\end{equation}
It follows from equation (\ref{eq:theta}) that they transform as 
\begin{equation}
\begin{aligned}
      \qty(O_8 ,V_8,S_8,C_8)(\tau+1)=&\qty(e^{-\frac{1}{3}\pi i}O_8,~-e^{-\frac{1}{3}\pi i}V_8,~e^{\frac{2}{3}\pi i}S_8,~e^{\frac{2}{3}\pi i}C_8)(\tau),\\
\begin{pmatrix}
    O_8 \\V_8 \\S_8 \\C_8
\end{pmatrix}\qty(-\frac{1}{\tau})=&\frac{1}{2}\left(\begin{array}{cccc}
1 & 1 & 1 & 1 \\
1 & 1 & -1 & -1 \\
1 & -1 & 1 & -1 \\
1 & -1 & -1 & 1
\end{array}\right)\begin{pmatrix}
    O_8 \\V_8 \\S_8 \\C_8
\end{pmatrix}(\tau).
\end{aligned}
\label{eq:fermions_modular_tr}\end{equation}
They have the following expansions:
\begin{equation}
    \begin{aligned}
        &O_8=\frac{1}{\eta^4}\qty(1+24q+\cdots),
        &&V_8=\frac{1}{\eta^4}\qty(8q^{\frac{1}{2}}+\cdots),\\
        &S_8=\frac{1}{\eta^4}\qty(8q^{\frac{1}{2}}+\cdots),
        &&C_8=\frac{1}{\eta^4}\qty(8q^{\frac{1}{2}}+\cdots).\\
    \end{aligned}
\end{equation}
The definition of the Dedekind eta function is:
\begin{equation}
    \eta(\tau)=q^{\frac{1}{24}}\prod_{n=1}^\infty \qty(1-q^n).
\label{eq:eta_function}\end{equation}
It has the following transformation properties:
\begin{equation}
    \begin{aligned}
        &\eta(\tau+1)=e^{\frac{\pi i}{12}}\eta(\tau),
        &&\eta\qty(-\frac{1}{\tau})=\sqrt{-i\tau}\eta(\tau).
    \end{aligned}
\label{eq:eta_modular_tr}\end{equation}

    \section{Root and Weight Lattices}\label{sec:lie algebra}
In this appendix, we summarize the details of the root lattices and weight lattices used in this paper. For notation and other details see \cite{Bourbaki:2002}.
We denote the $i$-th standard orthonormal basis of $\R^n$ by $\varepsilon_i$.
\subsection{\texorpdfstring{$A_n$}{An} type}
The root system of $A_n$ is
\begin{equation}
\begin{aligned}
    &\ve_j-\ve_k, 
    &&\text{for}\quad
    1\leq j<k\leq n+1, ~j\neq k.
\end{aligned}
\end{equation}
The basis are
        \begin{equation}
\begin{aligned}
        &\alpha_i=\varepsilon_i-\varepsilon_{i+1},&&\text{for}\quad 1\leq i \leq n.
\end{aligned}
\end{equation}
The fundamental weights of $A_n$ are
    \begin{equation}
        \begin{aligned}
            \omega_i=&\ve_1+\cdots+\ve_i-\frac{i}{n+1}(\ve_1+\cdots+\ve_{n+1})\\
            =&\frac{1}{n+1}\Bigl[(n-i+1)(\alpha_1+2\alpha_2+\cdots + (i-1)\alpha_{i-1})\\
            +&i\bigl((n-i+1)\alpha_i+(n-i)\alpha_{i+1}+\cdots+\alpha_n\bigr)\Bigr].
        \end{aligned}
    \end{equation}

\begin{equation}
    \begin{aligned}
        (\Lambda_R(A_n))^\ast=&\Lambda_W(A_n)\\
        =&\Lambda_R(A_n)+\Z\omega_1^{(A_n)},\\
    \Lambda_W(A_n)/\Lambda_R(A_n)=&\Z_{n+1}.
    \end{aligned}
\end{equation}

\subsection{\texorpdfstring{$C_n$}{Cn} type}
The root system of $C_n$ is
\begin{equation}
\begin{aligned}
     &\pm2\ve_j, 
     &&\text{for}\quad1\leq j \leq n,\\
     &\pm \ve_j\pm \ve_k, 
     &&\text{for}\quad1\leq j<k\leq n.
\end{aligned}
\end{equation}
The basis are
\begin{equation}
\begin{aligned}
        &\alpha_i=\varepsilon_i-\varepsilon_{i+1},&&\text{for}\quad1\leq i \leq n-1 ,\\
        &\alpha_n=2\ve_n.
\end{aligned}
\end{equation}
The fundamental weights of $C_n$ are
    \begin{equation}
\begin{aligned}
\omega_i= & \ve_1+\ve_2+\cdots+\ve_i \\
=&\alpha_1+2\alpha_2 +\cdots+(i-1) \alpha_{i-1} \\
+&i\left(\alpha_i+\alpha_{i+1}+\cdots+\alpha_{n-1}+\frac{1}{2} \alpha_n\right),
\end{aligned}
\end{equation}
The root lattice and weight lattice have following relationships:
\begin{equation}
    \begin{aligned}
        (\Lambda_R(C_n))^\ast=&\Lambda_W(C_n)+\Z\frac{1}{2}\omega_n\\
        =&\Lambda_R(C_n)+\Z\frac{1}{2}\alpha_n+\Z\frac{1}{2}\omega_n^{(C_n)}\\
    \Lambda_W(A_n)/&\Lambda_R(A_n)=\Z_{2}.
    \end{aligned}
\end{equation}

\subsubsection*{Quasi-Minuscule Representation}

In the root lattice of $C_n$, there are $2n(n-1)$ elements which have norm 2. By adding $n$ neutral elements, we obtain $(2n^2-n-1)$- and one-dimensional representation, where the former is called the quasi-minuscule representation of $C_n$.
\subsection{\texorpdfstring{$D_n$}{Dn} type}
The root system of $D_n$:
\begin{equation}
\begin{aligned}  &\pm\ve_j\pm\ve_k,\pm\ve_j\mp\ve_k,&&\text{for}\quad 1\leq j<k\leq n.
\end{aligned}
\end{equation}
The basis are
\begin{equation}
\begin{aligned}
        &\alpha_i=\ve_i-\ve_{i+1},&&\text{for}\quad 1\leq i\leq n-1, \\
        &\alpha_n=\ve_{n-1}+\ve_n.
\end{aligned}
\end{equation}
The fundamental weights of $D_n$ are   
\begin{equation}
\begin{aligned}
\omega_i= & \ve_1+\ve_2+\cdots+\ve_i\\
= & \alpha_1+2 \alpha_2+\cdots+(i-1) \alpha_{i-1}+i\left(\alpha_n+\alpha_{i+1}+\cdots+\alpha_{n-2}\right) \\
& +\frac{1}{2} i\left(\alpha_{n-1}+\alpha_n\right) , \quad\quad \text{for}\quad1 \leq i \leq n-2, \\
\omega_{n-1}= & \frac{1}{2}(\ve_n+\ve_2+\cdots+\ve_{n-2}+\ve_{n-1}-\ve_n) \\
= & \frac{1}{2}\left(\alpha_1+2 \alpha_2+\cdots+(n-2) \alpha_{n-2}+\frac{1}{2} n \alpha_{n-1}+\frac{1}{2}(n-2) \alpha_n\right) ,\\
\omega_n= & \frac{1}{2}\left(\ve_1+\ve_2+\cdots+\ve_{n-2}+\ve_{n-1}+\ve_n\right) \\
= & \frac{1}{2}\qty(\alpha_1+2 \alpha_2+\cdots+(n-2) \alpha_{n-2}+\frac{1}{2}(n-2) \alpha_{n-1}+\frac{1}{2} n\alpha_n).
\end{aligned}
\end{equation}
The root lattice and weight lattice have following relationships:

\begin{equation}
    \begin{aligned}
    \qty(\Lambda_R(D_{2n}))^\ast=&\Lambda_W(D_{2n})\\
    =&\Lambda_R(D_{2n})+\Z\omega_{2n-1}+\Z\omega_{2n},\\
    \qty(\Lambda_R(D_{2n+1}))^\ast=&\Lambda_W(D_{2n+1})\\
    =&\Lambda_R(D_{2n})+\Z\omega_{2n+1},\\
    \Lambda_{W}(D_{2n})/\Lambda_{R}(D_{2n})=\Z_2&\times\Z_2,
    \quad\quad\Lambda_{W}(D_{2n+1})/\Lambda_{R}(D_{2n+1})=\Z_4.
    \end{aligned}
\end{equation}

\subsection{\texorpdfstring{$E_6$}{E6}}
The root system of $E_6$ is
\begin{equation}
\begin{aligned}
    &\pm\varepsilon_i\pm\varepsilon_j, 
    &&\text{for}\quad1\leq i,j\leq 5,\\
    &\pm\frac{1}{2}\qty(\varepsilon_8-\varepsilon_7-\varepsilon_6+\sum_{i=1}^5 (-1)^{\nu_i}\varepsilon_i),
    &&\text{for}\quad\sum_{i=1}^5 \nu_i\in 2\Z.
\end{aligned}
\end{equation}
The basis are
\begin{equation}
\begin{aligned}
&\alpha_1=\frac{1}{2}\left(\varepsilon_1+\varepsilon_8\right)-\frac{1}{2}\left(\varepsilon_2+\varepsilon_3+\varepsilon_4+\varepsilon_5+\varepsilon_6+\varepsilon_7\right), &&\alpha_2=\varepsilon_1+\varepsilon_2, \\
& \alpha_3=\varepsilon_2-\varepsilon_1, &&\alpha_4=\varepsilon_3-\varepsilon_2, \\
&\alpha_5=\varepsilon_4-\varepsilon_3, &&\alpha_6=\varepsilon_5-\varepsilon_4.
\end{aligned}
\end{equation}
The fundamental weights of $E_6$ are
\begin{equation}
\begin{aligned}
\omega_1 =&\frac{2}{3}\left(\varepsilon_8-\varepsilon_7-\varepsilon_6\right)\\
=&\frac{1}{3}\left(4 \alpha_1+3 \alpha_2+5 \alpha_3+6 \alpha_4+4 \alpha_5+2 \alpha_6\right), \\
\omega_2 =&\frac{1}{2}\left(\varepsilon_1+\varepsilon_2+\varepsilon_3+\varepsilon_4+\varepsilon_5-\varepsilon_6-\varepsilon_7+\varepsilon_8\right) \\
=&\alpha_1+2 \alpha_2+2 \alpha_3+3 \alpha_4+2 \alpha_5+\alpha_6,\\
\omega_3  =&\frac{5}{6}\left(\varepsilon_8-\varepsilon_7-\varepsilon_6\right)+\frac{1}{2}\left(-\varepsilon_1+\varepsilon_2+\varepsilon_3+\varepsilon_4+\varepsilon_5\right) \\
=&\frac{1}{3}\left(5 \alpha_1+6 \alpha_2+10 \alpha_3+12 \alpha_4+8 \alpha_5+4 \alpha_6\right) ,\\
\omega_4 =&\varepsilon_3+\varepsilon_4+\varepsilon_5-\varepsilon_6-\varepsilon_7+\varepsilon_8 \\
=&2 \alpha_1+3 \alpha_2+4 \alpha_3+6 \alpha_4+4 \alpha_5+2 \alpha_6, \\
\omega_5 =&\frac{2}{3}\left(\varepsilon_8-\varepsilon_7-\varepsilon_6\right)+\varepsilon_4+\varepsilon_5 \\
=&\frac{1}{3}\left(4 \alpha_1+6 \alpha_2+8 \alpha_3+12 \alpha_4+10 \alpha_5+5 \alpha_6\right), \\
\omega_6 =&\frac{1}{3}\left(\varepsilon_8-\varepsilon_7-\varepsilon_6\right)+\varepsilon_5 \\ 
=&\frac{1}{3}\left(2 \alpha_1+3 \alpha_2+4 \alpha_3+6 \alpha_4+5 \alpha_5+4 \alpha_6\right) .
\end{aligned}
\end{equation}
The relation between the root and weight lattice is
\begin{equation}
    \begin{aligned}
    \qty(\Lambda_R(E_6))^\ast=&\Lambda_W(E_6)\\
    =&\Lambda_R(E_6)+\Z\omega_1 ,\\
    \Lambda_W(E_6)/\Lambda_R(E_6)=&\Z_3.
    \end{aligned}
\end{equation}

\subsection{\texorpdfstring{$E_7$}{E7}}
The root system of $E_7$ is
\begin{equation}
\begin{aligned}
    &\pm\varepsilon_i\pm\varepsilon_j, 
    &&\text{for}\quad1\leq i,j\leq 6,
    \\
    &\pm(\varepsilon_7-\varepsilon_8),\\
    &\pm\frac{1}{2}\qty(\varepsilon_8-\varepsilon_7-\varepsilon_6+\sum_{i=1}^6 (-1)^{\nu_i}\varepsilon_i),
    &&\text{for}\quad\sum_{i=1}^6 \nu_i\in 2\Z.
\end{aligned}
\end{equation}
The basis are
\begin{equation}
\begin{aligned}
& \alpha_1=\frac{1}{2}\left(\varepsilon_1+\varepsilon_8\right)-\frac{1}{2}\left(\varepsilon_2+\varepsilon_3+\varepsilon_4+\varepsilon_5+\varepsilon_6+\varepsilon_7\right), \\
& \alpha_2=\varepsilon_1+\varepsilon_2, \quad\quad\quad\quad\alpha_3=\varepsilon_2-\varepsilon_1, \\
& \alpha_4=\varepsilon_3-\varepsilon_2, \quad\quad\quad\quad \alpha_5=\varepsilon_4-\varepsilon_3, \\ 
& \alpha_6=\varepsilon_5-\varepsilon_4, \quad\quad\quad\quad \alpha_7=\varepsilon_6-\varepsilon_5 .
\end{aligned}
\end{equation}
The fundamental weights of $E_7$ are
\begin{equation}
\begin{aligned}
\omega_1 =&\varepsilon_8-\varepsilon_7\\
=&2 \alpha_1+2 \alpha_2+3 \alpha_3+4 \alpha_4+3 \alpha_5+2 \alpha_6+\alpha_7,\\
\omega_2 =&\frac{1}{2}\left(\varepsilon_1+\varepsilon_2+\varepsilon_3+\varepsilon_4+\varepsilon_5+\varepsilon_6-2 \varepsilon_7+2 \varepsilon_8\right) \\
=&\frac{1}{2}\left(4 \alpha_1+7 \alpha_2+8 \alpha_3+12 \alpha_4+9 \alpha_5+8 \alpha_6+3 \alpha_7\right), \\
\omega_3 =&\frac{1}{2}\left(-\varepsilon_1+\varepsilon_2+\varepsilon_3+\varepsilon_4+\varepsilon_5+\varepsilon_6-3 \varepsilon_7+3 \varepsilon_8\right) \\
=&3 \alpha_1+4 \alpha_2+6 \alpha_3+8 \alpha_4+6 \alpha_5+4 \alpha_6+2 \alpha_7, \\
\omega_4 =&\varepsilon_3+\varepsilon_4+\varepsilon_5+\varepsilon_6+2\left(\varepsilon_8-\varepsilon_7\right) \\
=&4 \alpha_1+6 \alpha_2+8 \alpha_3+12 \alpha_4+9 \alpha_5+6 \alpha_6+3 \alpha_7, \\
\omega_5 =&\varepsilon_4+\varepsilon_5+\varepsilon_6+\frac{3}{2}\left(\varepsilon_8-\varepsilon_7\right) \\
=&\frac{1}{2}\left(6 \alpha_1+9 \alpha_2+12 \alpha_3+18 \alpha_4+15 \alpha_5+10 \alpha_6+5 \alpha_7\right),\\
\omega_6 =&\varepsilon_5+\varepsilon_6-\varepsilon_7+\varepsilon_8 \\
=&2 \alpha_1+3 \alpha_2+4 \alpha_3+6 \alpha_4+5 \alpha_5+4 \alpha_6+2 \alpha_7, \\
\omega_7 =&\varepsilon_6+\frac{1}{2}\left(\varepsilon_8-\varepsilon_7\right) \\
=&\frac{1}{2}\left(2 \alpha_1+3 \alpha_2+4 \alpha_3+6 \alpha_4+5 \alpha_5+4 \alpha_6+3 \alpha_7\right) .
\end{aligned}
\end{equation}
The relation between the root and weight lattice is
\begin{equation}
    \begin{aligned}
    \qty(\Lambda_R(E_7))^\ast=&\Lambda_W(E_7)\\
    =&\Lambda_R(E_7)+\Z\omega_7,\\
    \Lambda_W(E_7)/\Lambda_R(E_7)=&\Z_2.
    \end{aligned}
\end{equation}

\subsection{\texorpdfstring{$E_8$}{E8}}

Note that for $E_8$, we use the same natation as in \cite{Font:2020rsk}.

The basis are

\begin{equation}
    \begin{aligned}
        \alpha_1=&\varepsilon_1-\varepsilon_2,\quad \quad \alpha_2=\varepsilon_2-\varepsilon_3,\\
        \alpha_3=&\varepsilon_3-\varepsilon_4,\quad \quad \alpha_4=\varepsilon_4-\varepsilon_5,\\
        \alpha_5=&\varepsilon_5-\varepsilon_6,\quad \quad \alpha_6=\varepsilon_6-\varepsilon_7,\\
        \alpha_7=&-(\varepsilon_1+\varepsilon_2),\quad\alpha_8=\frac{1}{2}\qty(\varepsilon_1+\cdots+\varepsilon_8).
    \end{aligned}
\end{equation}

\end{appendix}

\bibliographystyle{JHEP}
\bibliography{reference}

\end{document}